\documentstyle{article}

\title{Kinetics of aerosol formation.
1. Decay of metastable phase on several types of
heterogeneous centers}
\author{V.Kurasov}

\date{Victor.Kurasov@pobox.spbu.ru }

\begin{document}
\maketitle
\begin{abstract}

A system of a metastable phase with several sorts of the heterogeneous centers is
considered.
An analytical theory for the process of decay in such a system has been constructed.
The free energy of formation of the critical embryo  is assumed to be known in
the macroscopic approach as well as the  energy of solvatation.
The process is split into some periods and the analytical description of every
period is given. The most  difficult to describe is the
period of the essential formation of the embryos of a new phase. At first some
asymptotes are investigated and then a general solution is suggested.
Several approximate transformations are accomplished with the corresponding
numerical
estimates and some analytical justifications.
\end{abstract}

\pagebreak

\section{Introduction}

The case of   condensation of the supersaturated vapor into  the state
of  liquid droplets seems to be the simplest one among the first order
phase transitions. Traditionally this case  is treated as a model to
introduce some new theoretical  constructions in the description of the
first order phase transitions.  This leading role of the case  of
condensation was outlined by the creation  of the  classical theory of
 nucleation by Volmer \cite{1}, Becker and Doering \cite{2},
Zeldovitch
\cite{3} and Frenkel \cite{4} which gave for the first time a rather
simple expression for the rate of nucleation (i.e. for the rate of the
appearance
of new droplets).

The careful analysis of the classical  theory  leads to the great number
of the publications with the  various reconsiderations of the classical
expressions. Among them  one has to  notice the  account of the internal
degrees of freedom in the embryo made
by Lothe and Pound \cite{Lothe}, another modifications
made by Reiss, Cohen and Katz \cite{RCK}, Reiss \cite{Reiss}, Fisher \cite{Fisher}.
The application of the density functional theory  to the first order phase
transition made by R.Evans and D.Oxtoby \cite{EvOxt} and D.Oxtoby and
D.Zeng \cite{ZengOxt} allowed to put  the microscopic
(mesoscopic) models for the  condensating substance as the base for the
expression for the free energy of the critical embryo and for the rate
of nucleation.  The latest reconsideration of the classical theory can
be associated with  the contributions of Reiss, Tabazadech and Talbot
\cite{RTT}, Reiss, Ellerby and Weakliem \cite{REW1} - \cite{REW4},
Oxtoby and Talanquer \cite{OxtTal}.  In the cited publications the role
of the environment of the droplet is carefully analysed.  The choice
proposed in these publications
 is                           equivalent  to some specific  choice of
the statistical ensemble.

Nevertheless one has to notice that there is no perfect coincidence between
the concrete theoretical predictions\footnote{Certainly, one can not
check the general recipes as to calculate the objects like the statistical
sums.} and the experimental results. The macroscopic expression reproduces
only
the qualitative behavior of the experiment results. Meanwhile, the relative
deviation between  the  rate of
growth of the nucleation rate in the theoretical predictions
and in the experimental results is rather small.

These efforts  allowed to start the investigation of  kinetics
in  the field of the first order
phase transition.
The qualitative description of the phase transition was started
by Wakeshima \cite{Wak} who considered some time lags for condensation.
As far as the characteristic time of formation of the droplets spectrum
was necessary to ensure the correct experimental definition of the stationary
rate of nucleation namely this value was investigated in \cite{Wak}.
In \cite{Rai} the picture of the more realistic phase transition under
the smooth behavior     of the external conditions (i.e. under the external
conditions of the dynamic type) was given.
The form of the spectrum in these publications was introduced by some
artificial models.
But the real problem is to determine the form of the spectrum.

The homogeneous decay of a metastable phase was
investigated in \cite{9} by Kuni, Grinin, Kabanov. The analogous consideration
for the heterogeneous decay was fulfilled by Kuni in \cite{10} where only two
limit
situations were considered. The consideration of the simultaneous
 heterogeneous and
homogeneous decay
was made in \cite{11}, \cite{12} by the formal generalization of the
 procedure for the homogeneous decay \cite{9}.

The homogeneous condensation under the smooth behavior of the external
conditions (i.e. under the dynamic conditions) was investigated
in \cite{Kunidin}, \cite{univ}. The case of the
heterogeneous condensation was analyzed
in \cite{Novosib}, \cite{PhysRev} where the based qualitative descriptions
of the phase transition were given.

The external conditions of the dynamic type and the decay type are
the most natural external conditions and can be treated as some basic
ones.

Practically in
the condensing system there are
several types of the heterogeneous centers of a different nature.
Moreover in the process of condensation on the ions the free energy
of the critical embryo depends  on the sign of the electric charge.
As it is shown in \cite{14} the free energy $F$ of the near-critical embryo
in the state
of the internal
equilibrium
can be presented in the following manner:
\begin{equation} \label{1}
F=-b\nu+a\nu^{2/3}+c_{1}\nu^{1/3}+(c_{2}+c_{3})\nu^{-1/3} + c_{0} \ln \nu - G
\end{equation}
Here and in the further considerations all energy-like
values are expressed in the units of the thermal energy $k_{b}T$
($k_{b}$ is the Bolzman constant, $T$ is absolute temperature); $
 a, b, c_{0}, c_{1}, c_{2}, c_{3} $
are some constants; $G$ is the energy of the nuclei solvatation.
It necessary to notice that in contrast to $a, b, c_{0},
 c_{1}, c_{2}$ which don't depend
on the sign of the charge $q$ the value of $c_{3}$ is proportional to
$q$.
The value of $G$ also depends on the sign of $q$. This dependence is
similar to  dependence (\ref{1}) of $F$(without $G$).
 The only specific feature  we must fulfil
is to substitute the number of the molecules $\nu_{e}$ of the solvatated ion
instead of $\nu$.
As far as $\nu_{e} \neq \nu_{c}$ for the near-critical
embryo the value of $F$ depends on the sign
of $q$.
Hence, in the  presence of the
radiation we immediately have two sorts of the centers
(positive and negative) with the different heights of the activation barrier,
i.e. with the different activities of the heterogeneous centers.

When one has the spectrum of the sizes of the solid nucleus of
condensation with the weak interaction one must immediately come to the
spectrum of activities of the heterogeneous centers. Really, in the most
simple model appears when the nuclei is essentially noninteractive and one has
simply to add  the surface term to the number of the molecules imaginary
contained in the volume occupied by the nuclei. The free energy of the
critical embryo is given by
\begin{equation}
F = - b \nu + a
(\nu+\frac{4\pi r^3} {3 v_l})^{2/3} - G
\end{equation}
where $r$ is the
radius of the nuclei, $v_l$ is the volume occupied by a molecule of the
condensated substance in the liquid phase.  So, the spectrum of sizes of
the heterogeneous centers immediately initiates the spectrum       of the
activities of the heterogeneous centers\footnote{Later the value of the
activity  will be accurately defined}.

Any spectrum of the activities of the heterogeneous centers can be split
into  several continuous parts (may be also like $\delta$-functions) which
will be considered as the "types" of the heterogeneous centers.
So, one can see  four rather natural kinetic problems:
\begin{itemize}
             \item
The process of the decay
of the metastable state on the several types of the heterogeneous
centers.
\item
The process of the decay
of the metastabtable state on the continuous spectrum of the heterogeneous
centers.
  \item
The process of
condensation on the several types of the heterogeneous centers under the
dynamic conditions.
\item
The process of
condensation on the continuous spectrum of the heterogeneous centers under
the dynamic conditions.
\end{itemize}
The set of these publications is based on the ideas proposed in
\cite{density} - \cite{specdin}

Nevertheless the theory of the heterogeneous decay was constructed only
for one type of the heterogeneous centers. The task to construct the
 kinetic theory for these situations  is rather
essential. It will be completely fulfilled here.

The equations of condensation are rather similar
 in these four situations.
Unfortunately, the methods of their solution are
 absolutely different. This is
caused by the specific nonlinear character of the condensation equations.
Moreover, the analysis of any separate situation can not be spread on
another one. That's why the four separate publications are devoted to the
kinetics in these situations.

One has to study only the periods of the intensive formation of the droplets.
But  in contrast to condensation on the one sort of the heterogeneous
centers \cite{Novosib}, \cite{PhysRev} these periods aren't very  short\footnote{Under
the dynamic conditions}.

We shall use the following physical assumptions:
\begin{itemize}
\item
the thermodynamic description of the critical embryo,
\item
the random homogeneous space distribution of the
heterogeneous centers,
\item
the free-molecular regime of the droplets growth,
\item
the homogeneous
external conditions for the   temperature and for the pressure,
\item
rather a high
activation barrier\footnote{The theory without the heterogeneous activation
barrier is much more simple.},
\item
the absence of the thermal effects.
\end{itemize}

As far as according to  \cite{density} the condensation equations in the
general conditions of the condensation process are analogous to the case
of the free molecular consumption of the vapor we shall study namely this
case.

We assume the total number of the heterogeneous centers to be constant in
time.

As far as the most interesting characteristics of this process are the
 numbers of the heterogeneously
formed droplets of the different types
we shall estimate the accuracy of the theory
by
noticing the error of the obtained solutions for
 these values. The whole process of the metastable phase decay
can be split into two periods: the
period of the essential formation of the droplets  and
the period of the essential consumption of a metastable phase.
At first we shall
investigate the period of formation of the droplets.
 The unit volume is considered.
The consideration of the further evolution is rather simple and
can be  reduced
to the first order differential equation in the manner of  \cite{Kunidec}
for the situation of the decay and in the manner of \cite{Novosib}, \cite{PhysRev} for
the situation of the dynamic conditions.
We needn't to consider this period here.

In order to present  the clear descriptions some common facts will be
recalled at the beginning of every  publication. It allows to consider
the theories as the separate ones.

Below the situation of the decay of the metastable phase on the several
sorts of the heterogeneous centers will be described.

We assume the total  number of the heterogeneous centers to be constant
in time.

\section{Kinetic equation}

Suppose that there are several sorts of the heterogeneous centers. We
shall mark the total number of the heterogeneous centers by
$\eta_{tot\ i}$ where $i$ corresponds to some sort of the heterogeneous centers.
The real values of
the free heterogeneous centers which
may be solvatated but aren't occupied by
the super-critical embryos are marked by the value of $\eta_{i}$.
The index $i$ or
$j$ below the
value
marks the sort of the heterogeneous centers.
The sort of the droplets  means the sort of the heterogeneous centers.
The absence of this index points that
the formula is valid for an arbitrary sort of the heterogeneous centers.
          The density
of the molecules in the equilibrium vapor is marked
by \( n_{\infty} \), the
density
of the molecules in the real vapor is marked by the value \( n\).
 The power of the metastability will be characterized by the
value of the supersaturation
$$ \zeta = \frac{ n - n_{\infty} }{ n_{\infty} } $$
We shall define the super-critical embryos as the "droplets".
Every droplet is described by the molecule number
\( \nu \) , or by the linear size \( \rho = \nu^{1/3} \) .
Due to the free-molecular regime of the vapor consumption  we have
$$ \frac{d\rho}{dt} = \zeta \alpha \tau^{-1} $$
where \( \alpha \) is the condensation coefficient and \( \tau \) is
some characteristic
collision time obtained from the gas kinetic theory.

The characteristic time $t_*$ will be the time of the beginning of the
process of condensation. The values at the moment $t_*$ will be marked
by the lower index $t_*$.

The frontal type of the size spectrum allows to introduce the frontal size \( z \)
according to
\begin{equation}
\label{2}
z = \int_{0}^{t} \zeta \alpha \tau^{-1} dt'
\end{equation}
Until the coalescence \cite{15}, \cite{16}
which isn't considered here
equation (\ref{2}) ensures the growth of \( z \) in time and can be
inverted
\begin{equation} \label{3}
t(z) =\int_{0}^{z}  \tau \alpha^{-1} \frac{dx}{\zeta(x)}
\end{equation}
Hence, all values dependent on time become the values dependent on
\( z\) and the relative size \( x=z-\rho \) can be introduced.
During the whole evolution the droplet has one and the same
value of the variable \( x \).
Considering the value \( t(x) \) as the moment when the droplet with the
given $x$ has
been formed (as a droplet) we can consider the functions of
time as the functions
of \( x \) .
Hence, we can see that the kinetic equation is reduced to the
fact that every droplet keeps the constant value of $x$. To reconstruct the
picture
of the evolution one must establish the dependencies $t(z)$ and $\zeta(x)$.

\section{Condensation equations system}

We shall mark by the argument \( \infty \) the total values of the characteristics
formed during the whole condensation process.

Immediately after the creation  the value of the
supersaturation falls down to the value
\begin{equation} \label{4}
\Phi_{*} = \zeta (0) - \frac{
\sum_{i} \eta_{tot\ i}\nu_{e\ i}}{n_{\infty}}
\end{equation}
where \( \nu_{e\ i} \)
is the number of the
 molecules of the condensated substance
in the equilibrium heterogeneous embryo.
During the period of the essential formation of the droplets  one
can assume that the value \( \nu_{e\ i} \) is the constant one  and take
it at \( \zeta = \Phi_{*} \).
Ordinary \( \nu_{e\ i}\) can be taken at \( \zeta = \zeta(0) \).

The following statements can be analytically proved for the consideration
of this process:
\begin{itemize}
\item
(1) The main role in the vapor consumption during the evolution
is played by the super-critical embryos, i.e. by the droplets.
\item
(2) The quasistationary approximation for the nucleation rate is valid during the
period of the essential formation of the droplets.
\end{itemize}
The justification  of the second statement uses the estimate for the times
 \( t^{s}_{i} \)
of the establishing of the stationary  state
in the near-critical region  which can be found in \cite{3}, \cite{17}
(for the heterogeneous barrier  the consideration is  the same one).
Here it is necessary to remark that,
certainly, there may exists     some
rather huge times $t^{s}_{i}$. They correspond to some rather big values
of the number of the molecules inside the critical embryos.
The half-width of the near-critical region
can be estimated by the homogeneous value at the supersaturation corresponding
to the same  value of the
number of the molecules inside the critical embryo.
It is proportional to
$\nu^{2/3}_{c\ i}$. Here and in the further considerations the lower index "${c}$"
marks the values for the critical embryos.
As far as the absorption ability is
proportional to $\nu^{2/3}_{c\ i}$
and the size of the near-critical region is proportional to $\nu^{2/3}_{c}$ the
value of $t^{s}_{i}$
is proportional to $\nu^{2/3}_{c\ i}$.
The big value of the activation barrier $\Delta F_i = F_i(\nu_c) $
 in the case when
$\nu_{c\ i}$ is  greater than the characteristic
length $\Delta x$ of the size spectrum means that these
kinds of the heterogeneous centers are excluded
from the kinetic process.

One can analytically prove that
during the period of the intensive formation of the droplets of one sort
or for all sorts
\begin{itemize}
\item
the variation of the  supersaturation
allows the estimate
$$
\mid \zeta - \zeta_* \mid \leq \frac{\Phi_*}{\Gamma}
$$
\end{itemize}
where
\begin{equation}                   \label{6}
\Gamma_{i} = -\Phi_{*}
\frac{d \Delta_{i} F(\zeta)}{d \zeta }  \mid_{\zeta=\Phi_{*}}
\end{equation}

For the majority of the types of the heterogeneous centers
 the following approximations  of the nucleation rates \( J_{i} \)
 are valid during the period of the essential formation of the droplets
\begin{equation}          \label{5}
J_{i} = J_{i}( \eta_{tot\ i} , \Phi_{*} )
\exp ( \Gamma_{i} \frac{ ( \zeta - \Phi_{*} ) }
{ \Phi_{*} } )
\frac{\eta_i}
{\eta_{tot\ i}}
\end{equation}
The validity of these approximations is justified
for the heterogeneous embryos with
the monotonous interaction between the center and the molecules of the
condensated substance
 weaker or equal than the interaction reciprocal to the space distance.

Let \( f_{*\ i} \) be the amplitude value of the
distribution of sizes of the heterogeneously
 formed droplets
 measured in units of \( n_{\infty} \).
The supersaturation \( \Phi_{*} \) and the number of the heterogeneous
centers \( \eta_{tot\ i} \) are the already
known values. Then the stationary distribution $f$  can be easily
calculated by the following known formulas:
\begin{equation} \label{7}
\frac{J_i \tau}{\alpha \zeta n_{\infty}} =
f=\frac{ W^{+}_{c} \exp(-\Delta_i F) \tau}
{ n_{\infty} \pi^{1/2} \Delta_{e\ i} \nu  \Delta_{c\ i} \nu  \zeta  \alpha }
\eta_i
\end{equation}
where $W^{+}$ is the number of the molecules absorbed by the
critical embryo  in the unit
of time, $\Delta_{e} \nu$  is the characteristic width of
the equilibrium distribution
$$
\Delta_{e\ i} \nu = \sum_{\nu=1}^{\nu=(\nu_{c}+\nu_{e})/2}\exp(-F(\nu))
$$
and $\Delta_{c\ i} \nu$ is the halfwidth of the near-critical region
$$
\Delta_{c} \nu =
\frac{2^{1/2}}{\mid \frac{\partial^{2} F }{\partial
\nu^{2}}\mid^{1/2}_{\nu=\nu_{c}}}
$$

We shall mark by \( n_{\infty} g_{i} \) the total number of the vapor
molecules in the  droplets
formed on the centers of the sort "$i$".
To simplify the formulas we shall use $$ \theta_{i} =
\frac{\eta_{i}}{\eta_{tot\ i } } $$

Using the conservation laws for the  heterogeneous centers
and for the
  molecules
of the substance we obtain for \( g_{i},  \theta_{i} \) the following
equations
\begin{equation}\label{8}
g_{i} = f_{*\ i}  \int_{0}^{z} (z-x)^{3}
\exp ( -\Gamma_{i} \frac{ \sum_{j}g_{j}  }
{ \Phi_{*} } )
\theta_{i} dx
\equiv
G_{i}(\sum_{j}g_{j}, \theta_{i} )
\end{equation}
\begin{equation}\label{9}
\theta_{i} = \exp ( - f_{*\ i} \frac{n_{\infty}}{\eta_{tot\ i}} \int_{0}^{z}
\exp ( - \Gamma_{i} \frac{ \sum_{j}g_{j}  }
{ \Phi_{*} } ) dx )
\equiv
S_i( \sum_{j}g_{j})
\end{equation}
where $f_{*\ i} = J_{i}(\eta_{tot\ i},\Phi_{*})\tau /
\Phi_{*} \alpha n_{\infty}$.

These equations form the closed system of the condensation
 equations. This system will be the subject of our investigation.
For simplicity we shall investigate it for $i=1,2$.
As far as we measure the accuracy of the theory in the terms of
the error in the droplets number
 we  define these values as the following ones:
\begin{equation} \label{10}
N_{i} = \eta_{tot\ i} ( 1 - \theta_{i}(z)) \equiv Q_{i}(\theta_{i})
\end{equation}
The spectrum of sizes can be found as the following one
\begin{equation} \label{11}
f_{i}=
f_{*\ i}
\exp ( -\Gamma_{i} \frac{ \sum_{j}g_{j}  }
{ \Phi_{*} } )
\theta_{i}
\end{equation}

\section{Formal generalization of iteration method}

The formal transformation of the iteration method
for the process of condensation on the one sort of the heterogeneous
centers
leads to the following equations:
\begin{equation} \label{12}
g_{i\ (l+1)} = G_{i}(\sum_{j}g_{j\ (l)} , \theta_{i\ (l)})
\end{equation}
\begin{equation} \label{13}
\theta_{i \ (l+1)}=S_{i}(\sum_{j} g_{j\ (l)})
\end{equation}
\begin{equation} \label{14}
N_{i\ (l)} = Q_{i}(\theta_{i\ (l)})
\end{equation}
\begin{equation} \label{15}
g_{i\ (0)} = 0
\end{equation}
\begin{equation} \label{16}
\theta_{i\ (0)} = 1
\end{equation}
\begin{equation} \label{17}
g_{i\ (1)} = f_{*\ i} \frac{z^{4}}{4}
\end{equation}
\begin{equation} \label{18}
\theta_{i\ (1)} = \exp(-f_{*\ i} \frac{n_{\infty}}{\eta_{tot\ i}}
 z)
\end{equation}
\begin{equation} \label{19}
N_{i\ (2)}(\infty) = \eta_{tot \ i}
[ 1 -\exp(-f_{*\ i} \frac{n_{\infty}}{\eta_{tot\ i}}
(\sum_{j} \frac{\Gamma_{i}f_{*\ j}}{4\Phi_{*}})^{-1/4} A)]
\end{equation}
where
\begin{equation} \label{20}
A = \int_{0}^{\infty} \exp(-x^{4}) dx \approx 0.9
\end{equation}
The third iteration can not be calculated in the analytical form.
Meanwhile one can prove the convergence of the iterations analytically.

Let us analyze the expression for \( N_{i(2)}({\infty}) \). Assume
for a moment that for some $i$ and $j$
$$ f_{*\ i}   \gg f_{*\ j} $$
Let us decrease \( \eta_{tot\ i} \)
keeping the constant value of $ f_{*\ i} $
which is proportional to  $\eta_{i}$ with the fixed dependence on $\zeta$
by the increasing of the activity
of the heterogeneous centers of the sort "$i$",
i.e. by the decreasing of the height of the heterogeneous activation
barrier.
It is obvious that when $ \eta_{tot\ i} $  is small
then the total number of
the heterogeneously
formed droplets
coincides with
the total quantity of
the heterogeneous centers and goes to zero when $\eta_{tot\ i}$ goes to zero.
The value of $ g_{i}$ at
the end of the period of formation of the  droplets
on the heterogeneous centers of the sort "$j$" can be estimated as
$$
g_{i} \leq \frac{\eta_{tot\ i} (\hat{\Delta} x_{j})^{3}}{n_{\infty}}
$$
where $ \hat{\Delta} x_{j} $ is the width of the size spectrum (the size distribution
function) of  the droplets of the  sort "$j$".
The value of $ \hat{\Delta} x_{j}$ is above restricted by the
value $ \Delta x_{j} $ which is the width of the size
spectrum  without any  influence of
the droplets of the other sorts. Certainly, the value
of $\Delta x_j $
doesn't depend on $\eta_{tot\ i}$  and on $f_{*\ i}$ for all $i$.
 Hence, the influence
of the heterogeneous centers of the sort $i$ on the process of condensation
on the centers of the sort $j$ becomes
unessential (negligible) in the limit $ \eta_{tot\ i} \rightarrow 0 $.
At the same time the expression for $N_{j\ (2)}(\infty)$ shows that in the
limit
$ \eta_{tot\ i} \rightarrow 0 $, $ f_{*\ i}  = const $  the influence of
the heterogeneously  formed droplets of a sort $i$
doesn't become unessential. This leads to
the enormous  error in $N_{j}(\infty)$.
One cannot obtain the analytical expression in the third
approximation for $N_j$ in the  frames of the standard iteration method and
the second iteration gives the wrong qualitative results.

The reason for  the deviation in the results
 from the pure heterogeneous consideration is the following
one. In the case when the interruption
 of the embryos formation is caused by the
exhaustion of
the heterogeneous
centers the error in  the
value of $g$ is compensated by the squeezing force of the operator $S_i$. The
analogous property isn't valid for the operator $Q_i$
due to the cross influence of the droplets formed on some different sorts.

Note that  the procedure of this section can be successfully  applied
for the heterogeneous condensation on the centers of the one sort. Then
one can  analytically prove that  the relative error  of $N_{(2)}$ is
less than $0.015$.

\section{Limit cases}

The remarkable property of condensation in the situation
considered here is that
we can construct the theory by
the simple investigation
of all limit cases and cover  practically all situations.

At first we shall extract the characteristic sizes
in order to construct  rather a simple solution.

\subsection{Characteristic lengths}

The direct  generalization of the iteration method fails
due to the unappropriated account of the cross
influence of the heterogeneous droplets of some different sorts.
Nevertheless it allows
to define the spectrum of the droplets
in a proper  way when the cross influence is eliminated.
Hence, we
may use it to get the characteristics of the "self-formation" of
the droplets of the various sorts.

On a base of the first iterations in the  general procedure one can realize that
 for some process of a separate formation
there are two characteristic lengths. The first one is the
length of spectrum in the situation when
there are no exhaustion of the heterogeneous
centers (and no droplets of the other sort).
One can say that condensation occurs in the pseudo-homogeneous
way. For this characteristic value we have:
\begin{equation} \label{22}
\Delta_{i}x = (\frac{4\Phi_{*}}{\Gamma_{i} f_{*\ i}})^{1/4}
\end{equation}
This length is going from the first iteration for $g_{i}$ if the process
is split into  the separate  processes of condensation.
The second length is
the length of the spectrum
when there is no vapor exhaustion but only
the exhaustion of the heterogeneous centers. Then the width of
the spectrum is:
\begin{equation} \label{23}
\delta_{i}x = \frac{\eta_{tot\ i}}{f_{*\ i} n_{\infty} }
\end{equation}
This length is going from the first iteration for $\theta_{i}$.

Practically the hierarchy between $\Delta_{i}x, \delta_{j}x$ is ensured by
the hierarchy between
$f_{*\ i},\eta_{*\ j}$.  The values of $\Gamma_{i}$ are
rather (in comparison with $f_{*\ i}$) unsensible to the value of the
supersaturation. Really:
\begin{equation} \label{24}
-\frac {\Gamma_{i}}{\zeta} = \frac{d\Delta F}{d \zeta} \sim
 \frac{d F_{c}}{d \zeta}-  \frac{d G}{d \zeta}
\end{equation}
In the frameworks of the barrier character of condensation
($ \Delta F \gg 1$) we can give the above estimate for $(dF_{c} / d\zeta)$
for the  force of the interaction between the heterogeneous
center and the molecules of a liquid when this force  decreases monotonously in
 space.
The value of                             $(dF_{c} / d\zeta)$
can be above estimated
by  it's value in the limit of the homogeneous condensation
  $(dF_{c\ hom} / d\zeta)$.
As far as the energy of solvatation depends on the supersaturation
essentially
weaker than  $(dF_{c} / d\zeta)$ we can
neglect the last term of the previous equation and obtain:
\begin{equation} \label{25}
 \frac{d\Delta F}{d \zeta} \sim
 \frac{d F_{c\ hom}}{d \zeta}
\end{equation}
But this dependence is  rather weak one in comparison with the very sharp
$f_{*\ i}$ dependence on the supersaturation.

Another one important fact
 is the frontal character of the back side of the spectrum in the
pseudo-homogeneous situation (when $\Gamma_{i} $ plays some
role)\footnote{The
pseudo homogeneous situation can be treated  as the situation when the centers
of condensation remain practically unexhausted (in relative sense).}.
The frontal character can be seen from
\begin{equation} \label{26}
f_{i} = f_{*\ i} \exp(- \frac{\Gamma_{i}}{4 \Phi_{*}}(\sum_{j}f_{*\ j}) z^{4})
\end{equation}
Hence, the essential variation of the length $\Delta_{i}x$
can be caused only by the enormous variation
 of $f_{*\ i}$.

Instead of $\delta_{i} x $ we shall use the parameter
\begin{equation} \label{27}
h_{i} = \frac{\delta_{i}x}{\Delta_{i}x}
\end{equation}
In the reasons of simplicity we shall restrict ourselves by two sorts of
the heterogeneous centers.

\subsection{The case $\Delta_{1}x \sim \Delta _{2}x$}

\subsubsection{Situation $h_{1} \ll 1,\ \ h_{2} \geq 1$}

In this situation the following fact can be noticed:
\begin{itemize}
\item
The process of formation of the droplets on the heterogeneous centers
of the first sort (the "first sort droplets")  doesn't depend
on formation of the droplets of the second sort.
\end{itemize}
It can be
directly seen from the chain of inequalities
\begin{equation} \label{29}
\delta_{1} x \ll \Delta_{1} x \sim \Delta_{2} x \leq \delta_{2} x
\end{equation}
So we can describe the process of formation of the first sort droplets
 by the
following equalities:
\begin{equation} \label{30}
g_{1}=f_{*\ 1}
\int_{0}^{z} (z-x)^{3} \exp(-\Gamma_{1} \frac{g_{1}(x)}{\Phi_{*}}) \theta_{1}
dx
\equiv
G_{1}(g_{1},\theta_{1})
\end{equation}
\begin{equation} \label{31}
\theta_{1} = \exp(- f_{*\ 1} \frac{n_{\infty}}{\eta_{tot\ 1}}
\int_{0}^{z} \exp(-\Gamma_{1} \frac{g_{1}(x)}{\Phi_{*}})
dx )
\equiv
S_{1}(g_{1})
\end{equation}
 This system can be solved by the following iterations
\begin{equation} \label{32}
g_{1\ (i+1)} = G_{1}(g_{1\ (i)},\theta_{1\ (i)})
\end{equation}
\begin{equation} \label{33}
\theta_{1\ (i+1)}=S_{1}(g_{1\ (i)})
\end{equation}
\begin{equation} \label{34}
N_{1\ (i)} = Q_{1}(\theta_{1\ (i)})
\end{equation}
The operators $G_{1},S_{1}$ and $Q_{1}$ have some remarkable properties.
 When for all values of the arguments we have
$$ w_{1} \leq w_{2}$$
then
$$ S_{1}(w_{1}) \leq S_{1}(w_{2}) $$
is valid for all values of the arguments.
When  for all values of the arguments we have
$$ w_{1} \leq w_{2}$$
then
$$ Q_{1}(w_{1}) \geq Q_{1}(w_{2}) $$
is valid for all values of the arguments.
When  for all values of the arguments we have
$$ w_{1} \leq w_{2}$$
$$ v_{1} \geq v_{2}$$
then
$$ G_{1}(v_{1},w_{1}) \leq G_{1}(v_{2},w_{2}) $$
is valid for all values of the arguments.
Let us choose the initial approximations as the following ones
\begin{equation} \label{35}
g_{1\ (0)} = 0
\end{equation}
\begin{equation} \label{36}
\theta_{1\ (0)} = 1
\end{equation}
We see that
$$ g_{1\ (0)} \leq g_{1} $$
$$ g_{1\ (0)} \leq g_{1\ (i)} $$ for an arbitrary $i$, and
$$ \theta_{1\ (0)} \geq \theta $$
$$ \theta_{1\ (0)} \geq \theta_{1\ (i)} $$
for an arbitrary $i$. Particularly, the following estimate is
valid
\begin{equation}\label{37}
N_{1\ (2)} \leq N_{1} \leq N_{1\ (3)}
\end{equation}
These estimates allow to prove the convergence of the iterations.
The calculation of the iterations gives
\begin{equation} \label{38}
g_{1\ (1)} = f_{*\ 1} \frac{z^{4}}{4}
\end{equation}
\begin{equation} \label{39}
\theta_{1\ (1)} = \exp(-f_{*\ 1}
 \frac{n_{\infty}}{\eta_{tot\ 1}}z)
\end{equation}
\begin{equation} \label{40}
N_{1\ (2)}(\infty) = \eta_{tot\ 1} ( 1-\exp(-f_{*\ 1}
\frac{n_{\infty}}{\eta_{tot\ 1}}
 ( \frac{\Gamma_{1}}
{4 \Phi_{*}})^{-1/4}  f_{*\ 1}^{-1/4} A ))
\end{equation}
As far as one can prove that
\begin{equation} \label{41}
\frac{d}{dx} \mid N_{1\ (i)}-N_{1\ (j)} \mid  \geq 0
\end{equation}
then by
the simple numerical calculation of $ N_{1(3)}(\infty)$  one can easy obtain
that
  $$\frac{ \mid N_{1\ (2)} - N_{1} \mid }{N_{1}} \leq 0.015$$
which completes the iteration procedure.

On the base of the iterations one can get some approximations for the
supersaturation:
\begin{equation} \label{42}
\zeta_{(l+1)} = \Phi_{*} - f_{*\ 1} \int_{0}^{z} (z-x)^{3}
\exp(-\Gamma_{1}\frac{g_{1\ (l)}}{\Phi_{*}}) \theta_{1\ (l)} dx
\end{equation}
The strong inequality allows to obtain
the second approximation for $\zeta$
\begin{equation} \label{43}
\zeta_{(2)} = \Phi_{*} - f_{*\ 1} \int_{0}^{z} (z-x)^{3}
\exp(-H x) dx
\end{equation}
where
\begin{equation} \label{44}
H = \frac{f_{*\ 1} n_{\infty}}{\eta_{tot\ 1}}
\end{equation}
or after the calculation
\begin{equation} \label{45}
\zeta_{(2)} = \Phi_{*}+f_{*\ 1}[ -
\frac{z^{3}}{H} +
\frac{3 z^{2}}{H^2}-
\frac{6 z}{H^3}+
\frac{6}{H^4}-
\frac{6}{H^4}\exp(-Hz)]
\end{equation}

This expression can be simplified. Let us notice that the supersaturation
appears in the expression for the size spectrum $f(x)$ in the
following form
$$ \exp(-\Gamma_{i} \frac{\zeta-\Phi_{*}}{\Phi_{*}}) $$

After the substitution of $\zeta_{(2)}$ into this expression we
realize that in the case when $\zeta$  deviates  essentially from $\Phi$
all the terms except the first two terms can be neglected:
\begin{equation} \label{46}
\zeta_{(2)} = \Phi_{*} - z^3 \frac{\eta_{tot\ 1}}{n_{\infty}}
\end{equation}

Hence, for the second sort we can obtain the following system of equations
\begin{equation} \label{47}
g_{2}=f_{*\ 2}
\int_{0}^{z} (z-x)^{3} \exp(-\Gamma_{2} \frac{g_{2}(x)
+(\eta_{tot\ 1}/n_{\infty})x^3
}{\Phi_{*}}) \theta_{2} dx
\equiv
G_{2}(g_{2}
+(\eta_{tot\ 1}/n_{\infty})x^3,\theta_{2})
\end{equation}
and
\begin{equation} \label{48}
\theta_{2} = \exp(- f_{*\ 2} \frac{n_{\infty}}{\eta_{tot\ 2}}
\int_{0}^{z} \exp(-\Gamma_{2} \frac{g_{2}(x)
+(\eta_{tot\ 1}/n_{\infty})x^3
}{\Phi_{*}})
dx )
\equiv
S_{2}(g_{2}
+(\eta_{tot\ 1}/n_{\infty})x^3)
\end{equation}

Having introduced
\begin{equation} \label{49}
\lambda_{2}=g_{2}
+(\eta_{tot\ 1}/n_{\infty})z^3
\end{equation}
we can rewrite this system as
\begin{equation} \label{50}
\lambda_{2} = G_{2}(\lambda_{2},\theta_{2})
+(\eta_{tot\ 1}/n_{\infty})z^3
\equiv
G_{2}^{+}(\lambda_{2},\theta_{2})
\end{equation}
\begin{equation} \label{51}
\theta_{2} = S_{2}(\lambda_{2})
\end{equation}

The operator $G^{+}_{2}$ has the same properties as $G_{1},G_{2}$ have.
All estimates (\ref{35})-(\ref{37})
remain valid after the substitution of the index "$2$"
instead of "$1$" and the operator $G^{+}$ instead of $G$.
Moreover one can see that
\begin{equation} \label{52}
\frac{d}{d(\eta_{tot\ i}/n_{\infty})} \mid N_{2\ (i)} - N_{2\ (j)} \mid \leq 0
\end{equation}
which leads to
the same estimate of the relative
error  (0.015) as in the pure heterogeneous condensation of
a separate sort.

Actually  we can avoid here the
calculations according to  such a complex procedure. Let
us notice that the term $\eta_{tot\ 1} z^3/ n_{\infty}$ ensures
the characteristic length
\begin{equation} \label{53}
D_{1}=(\frac{\Phi_{*} n_{\infty}}{\Gamma_{2} \eta_{tot\ 1}})^{1/3}
\end{equation}
Really
\begin{equation} \label{54}
D_{1} \geq \epsilon \Delta_{1} x \sim \epsilon \Delta_{2} x \ \ \ \ \
\epsilon\sim (3 \pm 1 )
\end{equation}
So, condensation on the centers of the second sort occurs in the separate
way and we can use   formulas (\ref{30})-(\ref{34}), (\ref{38})-(\ref{41})
 with the substitution of the
index "$2$" instead of the index "$1$".

\subsubsection{Situation $h_{1} \geq 1,\ \ \ h_{2}\ll 1$}

As far as $\Delta_{1} \sim \Delta_{2}$ we can change the numbers of sorts and
reduce this situation to the previous one.

\subsubsection{Situation $h_{1} \geq 1,\ \ \ h_{2} \geq 1$}

In order to analyze this situation we must realize why in the separate
condensation of one sort the second iteration gives
rather precise results.
This fact is explained by the free molecular regime of the vapor consumption
 which
leads to the power $3$ in
the expression for $g$. Due to rather
a great (in comparison with 1) value of this power the droplets
of the big sizes situated
near the front side  of the spectrum are the main consumers of the
vapor. These droplets have rather a
small (in comparison with $\Delta_{i}x$) values of the variable $x$.
The exhaustion of
the heterogeneous centers due to $h_{1} \geq 1, h_{2} \geq1 $
doesn't affect in a strong manner on the process of
their formation. Hence, the cross influence is
rather weak and we can use the general iteration
procedure. After the calculation of the iterations we have
\begin{equation} \label{55}
g_{1\ (1)} = f_{*\ 1} \frac{z^{4}}{4}
\end{equation}
\begin{equation} \label{56}
\theta_{1\ (1)} = \exp(-f_{*\ 1} \frac{n_{\infty}}{\eta_{tot\ 1}}
 z)
\end{equation}
\begin{equation} \label{57}
N_{1\ (2)}(\infty) = \eta_{tot \ 1}
[ 1 -\exp(-f_{*\ 1} \frac{n_{\infty}}{\eta_{tot\ 1}}
( (\frac{\Gamma_{1}}{4\Phi_{*}})f_{*\ 1}+
(\frac{\Gamma_{1}}{4\Phi_{*}})f_{*\ 2})^{-1/4}A)]
\end{equation}
\begin{equation} \label{58}
g_{2\ (1)} = f_{*\ 2} \frac{z^{4}}{4}
\end{equation}
\begin{equation} \label{59}
\theta_{2\ (1)} = \exp(-f_{*\ 2} \frac{n_{\infty}}{\eta_{tot\ 2}}
 z)
\end{equation}
\begin{equation} \label{60}
N_{2\ (2)}(\infty) = \eta_{tot \ 2}
[ 1 -\exp(-f_{*\ 2} \frac{n_{\infty}}{\eta_{tot\ 2}}
( (\frac{\Gamma_{2}}{4\Phi_{*}})f_{*\ 1}+
(\frac{\Gamma_{2}}{4\Phi_{*}})f_{*\ 2})^{-1/4}A)]
\end{equation}

\subsubsection{Situation $h_{1}\ll 1, h_{2} \ll 1$}

As far as
\begin{equation} \label{61}
\delta_{1}x \ll \Delta_{1}x \sim \Delta_{2}x
\end{equation}
the droplets formed on the heterogeneous centers of the
second sort don't act on the pricess of formation  of
the droplets on the heterogeneous centers of the first sort.
Due to
\begin{equation} \label{62}
\delta_{2} \ll \Delta_{2}x \sim \Delta_{1}x
\end{equation}
the same fact is
valid for the droplets formed   on  the other sort of the heterogeneous
centers.
Hence, the system is split into some parts corresponding to the
separate processes of condensation on the different sorts.
Constructions (\ref{30})-(\ref{34}) can be reproduced here.
But in the case $h_{i} \ll 1$
for all $i$ the vapor exhaustion can be neglected in comparison
with the exhaustion of the heterogeneous centers.

In this situation we can obtain some
precise explicit results for the values of
$f_{i}(x), \zeta(x), \theta_{i}(x)$:
\begin{equation}\label{63}
\theta_{1}(x) = \exp(-f_{*\ 1} \frac{n_{\infty}}{\eta_{tot\ 1}} z)
\end{equation}
\begin{equation} \label{64}
\theta_{2}(x) = \exp(-f_{*\ 2} \frac{n_{\infty}}{\eta_{tot\ 2}} z)
\end{equation}
\begin{equation}\label{65}
f_{1}(x) = f_{*} \exp(-f_{*\ 1} \frac{n_{\infty}}{\eta_{tot\ 1}} z)
\end{equation}
\begin{equation} \label{66}
f_{2}(x) = f_{*} \exp(-f_{*\ 2} \frac{n_{\infty}}{\eta_{tot\ 2}} z)
\end{equation}
\begin{equation} \label{67}
\zeta = \Phi_{*} - (\frac{\eta_{tot\ 1}}{n_{\infty}} +
\frac{\eta_{tot\ 2}}{n_{\infty}}) z^3
\end{equation}
This expression for $\zeta$ is obtained by the same procedure
as that which led to (\ref{46}).

\subsection{Case $\Delta_{1} x \ll \Delta_{2} x$}

The case
$\Delta_2 x  \gg \Delta_1 x $
is simmetrical to this case and can be considered by the simple change
of the indexes.

Due to $\Delta_{1} x \ll \Delta_{2} x$
the droplets of the second sort don't
act on the process of formation of the droplets of the first sort.
Hence, the process of
formation of the droplets of the first sort can be described by
the iteration procedure from the  section 5.2.1 (eq.(\ref{30})-(\ref{41})).

\subsubsection{Situation $h_{1}\ll 1, h_{2} \geq 1 $}

Due to $h_{1}\ll 1$ the equation for the first sort can be
simplified and we have the following equations
\begin{equation} \label{68}
g_{1}= f_{*\ 1} \int_{0}^{z} (z-x)^3 \exp(-Hx) dx \sim
\frac{\eta_{tot \ 1}}{n_{\infty}} z^3
\end{equation}
The value of $\theta_{1}$ is given by (\ref{63}),
 the value of $f_{1}$ is given by (\ref{65}).

For condensation on the centers of the
second sort we have some equations analogous to (\ref{47}),(\ref{48}).
So we can adopt here equations (\ref{49})-(\ref{52}).
But in this situation  inequality (\ref{54})
isn't valid and we must calculate the iterations.
We can choose $\lambda_{2\ (0)} = 0$ and get
\begin{equation} \label{69}
\lambda_{2\ (1)} = f_{*\ 2}\frac{z^{4}}{4} +
\frac{\eta_{tot\ 1}}{n_{\infty}} z^{3}
\end{equation}
\begin{equation} \label{70}
\theta_{2\ (2)}=
\exp(- f_{*\ 2}
\frac{n_{\infty}}{\eta_{tot\ 2}} \int_{0}^{z}
\exp(-( \frac{x}{\Delta_{\infty\ 2}x})^{4} - (\frac{x}{\Delta_{h\ 1} x })^{3})
dx)
\end{equation}
where
$$ \Delta_{\infty \ 2} x = (\frac{4 \Phi_{*}}{\Gamma_{2} f_{*\ 2}})^{1/4}
\equiv
\Delta_2 x
$$
and
$$ \Delta_{h \ 1} x = ( \frac{\Phi_{*}n_{\infty}}
{\Gamma_{2} \eta_{tot\ 1}} )^{1/3}
\equiv
D_1
$$

The value of $\Delta_{\infty \ 2} x$ has
the sense of the spectrum width when the
cross influence and the  exhaustion  of the heterogeneous
centers                            of this very sort
are neglected. The value of $\Delta_{h \ 1} x $
 has the sense of the spectrum width
when the vapor consumption  by the droplets is neglected.

In addition one can easily prove that
$$ \frac{d}{dx} \mid N_{2\ (i)} - N_{2\ (j)} \mid \geq 0 $$
and
$$ \frac{d}{d\eta_{tot\ 1}} \mid N_{2\ (i)} - N_{2\ (j)} \mid \leq 0 $$
for $i,j \geq 2$.
Hence, it is easy to show that
$$ \frac{\mid N_{2\ (2)} - N_{2} \mid}{N_{2}} \leq 0.015 $$
by the calculation of  $N_{2\ (2)}(\infty)$ and
$N_{2\ (3)}(\infty)$ at $\eta_{tot\ 1} = 0 $

The analytical approximation valid for the clear interpretation can be
obtained
if we notice that
\begin{equation} \label{71}
\theta_{2\ (2)}(\infty) = \exp[-f_{*\ 2} \Delta_{\infty \ 2} x
\frac{n_{\infty}}{\eta_{tot\ 2}}
(\frac{A}{2}(1+ (\frac{\Delta_{\infty \ 2}x}{\Delta_{h \ 1}x})^{4})^{-1/4} +
\frac{B}{2}(1+ (\frac{\Delta_{\infty \ 2 }x}{\Delta_{h\ 1}x})^{3})^{-1/3})]
\end{equation}
where
$$ B = \int_{0}^{\infty}\exp(-x^{3})
dy $$
with the relative error less than 0.02.

The spectrum of sizes of the droplets formed on the centers
of the second sort is the following one
\begin{equation} \label{72}
f_{2} = f_{*\ 2} \exp(-\frac{\Gamma_{2} f_{*\ 2}}{\Phi_{*}} \frac{z^4}{4})
\exp(-\frac{\Gamma_2}{\Phi_*} \frac{\eta_{tot 1}}{n_{\infty}} z^3)
\exp( - f_{*\ 2}\frac{n_{\infty}}{\eta_{tot\ 2}} \int_{0}^{z}
\exp( - (\frac{x}{\Delta_{\infty \ 2 }x})^{4} - (\frac{x}{\Delta_{h \ 1} x
})^{3}) dx)
\end{equation}

\subsubsection{Situation $h_{1} \geq 1, h_{2} \ll 1$}

The description of the process of formation
of the droplets on the centers of the first sort  can't be simplified.
It has been already fulfilled in the previous sections.
But the process of formation of the droplets on the centers of the second
sort
 is rather simple to describe.
The supersaturation is absolutely
determined  by  the vapor consumption  by the droplets
formed on the centers of the first sort. Hence, we have the following expressions
\begin{equation} \label{73}
\theta_{2} = \exp ( - f_{*\ 2} \frac{n_{\infty}}{\eta_{tot\ 2}} \int_{0}^{z}
\exp ( - \Gamma_{2} \frac{ g_{1}  }
{ \Phi_{*} } ) dx )
\end{equation}
\begin{equation} \label{74}
g_{2} = f_{*\ 2}  \int_{0}^{z} (z-x)^{3}
\exp ( -\Gamma_{2} \frac{ g_{1}  }
{ \Phi_{*} } )
\theta_{2} dx
\end{equation}

The value of  $g_2$ during the period of  nucleation on the centers
of the second sort can be estimated in the following way
$$
g_2 \ll \frac{\Phi_*}{\Gamma_2}
$$
which is presented  on the base of
$$
\delta_2 x  \ll \Delta_2 x
$$
It is necessary to calculate only $\theta_2$. To calculate $\theta_2$
one can get into account that the value of  $g_1$ grows so rapidly that
for the value
$ \int_0^z \exp(-\Gamma_2 \frac{g_1}{\Phi_*} )dx
$
one can show  the following approximation
$$
  \int_0^z \exp(-\Gamma_2 \frac{g_1}{\Phi_*}) dx
\approx
z \Theta(1-\frac{\Gamma_2 g_1}{\Phi_*}) +
\int_0^{\infty} \exp(-\Gamma_2 \frac{g_1}{\Phi_*}) dx
\Theta(\frac{\Gamma_2 g_1}{\Phi_*} - 1)
$$
$$
  \int_0^z \exp(-\Gamma_2 \frac{g_1}{\Phi_*}) dx
\approx
z \Theta(1-\frac{\Gamma_2 g_1}{\Phi_*}) +
z_b \Theta(\frac{\Gamma_2 g_1}{\Phi_*} - 1)
$$
where $z_b$ is extracted by the condition
$$ g_1(z_b) = \frac{\Phi_*}{\Gamma_2} $$
and $\Theta$ is the Heavisaid's function.
The last approximation solves the problem of the analytical calculation of
$\theta_2$.

\subsubsection{Situation $h_{1} \geq 1, h_{2} \geq 1$ }

Actually, the situation $h_{1} \geq 1, h_{2} \geq 1$
has been already described in the previous subsection. Due
to $\Delta_{1}x \ll \Delta_{2}x$
one can't
assume here that the inequality $h_{2} \ll 1$ ensures  the
pure exhaustion of the heterogeneous centers without any vapor exhaustion
and the consideration
made in the previous section can't be simplified.

\subsubsection{Situation $h_{1} \ll 1, h_{2} \ll 1$ }

From the first point of view it seems that
the situation $h_{1} \ll 1, h_{2} \ll 1$ has been  already described in
the section 5.2.4.
We have to stress that the equation $h \ll 1$ doesn't allow to state that the
process of condensation is going at the constant value of the supersaturation.
For the process of condensation on the centers of the first sort we
have  previous expressions (\ref{63}), (\ref{65}). Analogous expressions
(\ref{64}), (\ref{66}) for condensation on the centers of the
 second sort can be violated.                                 So, the
process of condensation on the centers of the second sort can not
be described on the base of the unexhausted value of the vapor supersaturation.

The calculation  of $g_{2}$ isn't necessary and only the calculation of
$\theta_{2}
$ is essential. We have
\begin{equation} \label{75}
\theta_{2\ (2)}= \exp(- f_{*\ 2}\frac{n_{\infty}}{\eta_{tot\ 2}} \int_{0}^{z}
\exp(( - (\frac{x}{\Delta_{h \ 1} x })^{3}) dx)
\end{equation}
and the final value for $\theta_{2}$ can be given by
\begin{equation} \label{76}
\theta_{2\ (2)}(\infty) = \exp[-f_{*\ 2} \Delta_{h\ 1} x
\frac{n_{\infty}}{\eta_{tot\ 2}}B]
\end{equation}

The monodisperse approximation is based on the evident chain
of the inequalities
$$
\hat{\Delta} x_1 \sim \delta_1 x \ll \Delta_1 x \leq \hat{\Delta} x_2
$$
or the monodisperse term isn't essential.

We have investigated all  possible situations.\footnote{It doesn't follow that
all of them can be really reproduced in the nature.}

\section{General approximate solution}

The reason why the general iteration procedure fails lies in the fact that we
don't know the true expression for $g_{i}(z)$. In the case of
condensation on the centers of the
separate sort  we don't know  this expression only
when the spectrum is cut off by the exhaustion of the heterogeneous centers.
 So we don't know it in the situation then the converging force of
the operator in the expression for $\theta$ is very strong. In
the case of the several components
the situation is
different. We don't know every term in $\sum_{j} g_{j}(z)$.
Hence,
we may come to the situation when according  to the first iteration the spectrum
is cut off by the exhaustion of the supersaturation initiated by the other component but
in reality
 the heterogeneous centers of the other component are exhausted and
and the droplets formed on these centers consume the vapor much more
weaker (simply due to their quantity).
So, it is necessary to invent the new more precise expression for
$g_{i}$ which allows to calculate the next iteration in $\theta$.

\subsection{Monodisperse approximation}

As it is stated in section 5.1 the length
corresponding to the cut-off  of  the supersaturation
which leads to the cut-off of the  spectrum by the exhaustion of
the substance is practically one and the same for all sorts of the droplets (all
$\Gamma_i$ have approximately one and the same order).
Let us see the droplets of what sizes play the leading role in this cut-off.
Analyzing the subintergal expression in the equation for  $g_{i}$
we realize that the
 subintegral expression connected with the variation of the supersaturation
is  a very sharp function of $x$. It is less than the function
\begin{equation} \label{77}
s_{bel} = \Theta(z-x) (z-x)^3
\end{equation}
and greater than the function
\begin{equation} \label{78}
s_{ab} = \Theta(z-x) (z-x)^3 \exp(-\frac{\Gamma_{i}\sum_j f_{* j}}
{\Phi_{*}} \frac{x^4}{4} )
\end{equation}

Let us introduce the approximation for this function. We must
extract the region of the sizes of the droplets  which are essential in
the vapor consumption.
This consumption  is essential when
\begin{equation} \label{79}
x
\approx
\Delta x
\end{equation}
where $\Delta x$ is the length of the cut-off by the supersaturation.
Certainly,
the region of the sizes of the droplets  which are essential in
the vapor consumption must have the sizes rather small in comparison with
$\Delta_{i}x$ because the successful
 iteration procedure in the homogeneous  decay
is based on the fact
that the droplets formed at the almost ideal supersaturation
determine  the process
of formation of the spectrum.
For the differential halfwidth $\delta_{1/2}$ we have the following expression
\begin{equation} \label{80}
\delta_{1/2}x = ( 1 - \frac{1}{2^{1/3}} ) x
\end{equation}
The integral halfwidth $\Delta_{1/2}x$ can be obtained from the
corresponding equation
\begin{equation} \label{81}
N_{ess}x^3 = f_{*\ i}\frac{x^4}{4} n_{\infty}
\end{equation}
where $N_{ess} $ is the characteristic number of the  droplets obtained as
$N_{ess}=f_{*\ i}\Delta_{1/2}x n_{\infty}$ which gives
\begin{equation} \label{82}
\Delta_{1/2}x = \frac{1}{4} x
\end{equation}
and it practically coincides with $\delta_{1/2}x$.

Figure 1 illustrates   the behavior of the values $s_{ab}$ and
$s_{bel}$ as the functions of $x$. It can be seen that they practically
coincide. This lies in the base of the applicability of the first iteration
as a good approximation for the precise solution.

So, the subintegral function $s$ is now decomposed into
the essential part where
$$x \leq \frac{\Delta_{i}x}{4}$$ and the tail where $$x \geq
 \frac{\Delta_{i}x}{4}$$
We shall neglect the tail and  due to rather a small size of the essential region
we shall
use the monodisperse approximation for the droplets formed in this region.
As the result we obtain the approximation for $g(x)$
\begin{equation} \label{83}
g(z) = \frac{N(z/4)}{n_{\infty}}z^3
\end{equation}
where $N(z/4)$ is the number of the droplets appeared from $x=0$ till $x=z/4$.

\begin{it}
As far as
the spectrum is  cut off by the exhaustion of the supersaturation  in
a frontal (sharp) manner
 the value of $g_i$ is unessential
before $z=\Delta_{i}x$
as a small one and after the
moment of the cut-off
it is unessential also as there is no formation of the droplets.
\end{it}

So
instead of the previous approximation we can use
\begin{equation} \label{84}
g_{i}(z) = \frac{N_{i}(\Delta_{i} x/4)}{n_{\infty}}z^3
\end{equation}
The process of the exhaustion of
the heterogeneous centers makes the subintegral function
more sharp and the monodisperse approximation  becomes  at
$\Delta_{i}x$ even better than in the pseudo homogeneous situation.
 But the exhaustion of the heterogeneous centers makes the
coordinate of the cut-off of the supersaturation  greater than $\Delta_{i}x$ and
the monodisperse approximation becomes even more better at
the moment of the cut off of the imaginary supersaturation.
 Certainly, we must use $ N(\Delta_{i} x /4)$
calculated with account of
the exhaustion of the heterogeneous centers (but at the coordinate, obtained without
any account of the exhaustion of the heterogeneous centers).

Figure 2 illustrates the form of the spectrum in the "monodisperse approximation".
The case of the pseudo homogeneous condensation is considered.
Here two curves are drawn: the spectrum in the monodisperse approximation
$f_{appr}$
and the spectrum in the first iteration $f_{1}$ which can
be considered as a very precise approximation. The first iteration has
the more sharp back side (the front side in $z$ scale) than the "monodisperse
approximation.  Nevertheless the deviation isn't so essential.
It can be eliminated by the perturbation theory.

The concluding remarks concern the fact that
we can  obtain $ N(\Delta_{i}x/4)$
by the solution of the equations for the
separate condensation process  because
we need the lowest length of the cut-off. This length is given without
any cross influence
taking into account due to the frontal character of the back side of
the spectrum.

\subsection{Final iterations}

The solution of the system of the condensation equations
is given by the following procedure.

At first we must solve the equations for the separate processes
\begin{equation} \label{85}
g_{i} = f_{*\ i}  \int_{0}^{z} (z-x)^{3}
\exp ( -\Gamma_{i} \frac{ g_{i}  }
{ \Phi_{*} } )
\theta_{i} dx
\equiv
G_{i}(g_{i}, \theta_{i} )
\end{equation}

\begin{equation} \label{86}
\theta_{i} = \exp ( - f_{*\ i} \frac{n_{\infty}}{\eta_{tot\ i}} \int_{0}^{z}
\exp ( - \Gamma_{i} \frac{ g_{i}  }
{ \Phi_{*} } ) dx )
\equiv
S_i( g_{i})
\end{equation}
for every $i$.

This solution can be obtained by the iteration procedure described in section 4.
The remarkable fact is that in the second approximation
we can calculate the value of $\theta_{(2)}(\Delta_{i}x/4)$
\begin{equation} \label{87}
\theta_{i (2)}(\Delta_{i}x/4) =
\exp(- f_{*\ i} \frac{n_{\infty}}
{\eta_{tot\ i}} \int_{0}^{\Delta_{i}x/4}
\exp(-\frac{\Gamma_{i}f_{*\ i}}{4 \Phi_{*}} z^4) dx )
\end{equation}

\begin{equation} \label{88}
\theta_{i (2)}(\Delta_{i}x/4) =
\exp(- f_{*\ i} \frac{n_{\infty}}{\eta_{tot\ i}}
(\frac{\Gamma_{i}f_{*\ i}}{4 \Phi_{*}})^{-1/4} C )
\end{equation}
where
\begin{equation} \label{89}
C= \int_{0}^{1/4} \exp(-z^4) dx \approx 0.25
\end{equation}
and $N_{i (2)}(\Delta_{i}x/4)$ has the following value
\begin{equation} \label{90}
N_{i (2)}(\Delta_{i}x/4) = \eta_{tot\ i}(1-\theta_{i (2)}(\Delta_{i}x/4))
\end{equation}
So the approximations for $g_{i}$ are obtained now. We must fulfil
these calculations for every sort of the heterogeneous centers.
Considering these approximations as the initial ones it is necessary to
do only one step of the iteration procedure to get the suitable results.
They will be marked as the "final" ones.

Now we remove to the general iteration procedure. The only thing we have to do
is to calculate $\theta_{i\ final}(\infty)$
and $N_{final\ i}(\infty)$. We have to obtain these values
due to
the iterations
\begin{equation} \label{91}
\theta_{i\ final}(z) =
\exp[- f_{*\ i} \frac{n_{\infty}}{\eta_{tot\ i}} \int_{0}^{z}
\exp(-
\frac{\sum_{j}\Gamma_{i}N_{j\ (2)}(\Delta_{j}x/4)}
{ n_{\infty}\Phi_{*}} z^3) dx]
\end{equation}
\begin{equation} \label{92}
N_{final\ i}(\infty) = \eta_{tot\ i}[
1-
\exp(- f_{*\ i} \frac{n_{\infty}}{\eta_{tot\ i}}
(\frac{\sum_{j}\Gamma_{i}N_{j\ (2)}(\Delta_{j}x/4)}
{ n_{\infty}\Phi_{*}})^{-1/3} B
]
\end{equation}

All these equations are valid under the reasonable separation of the heterogeneous
centers into the sorts when the centers with one and the same height of
the activation barrier are considered as one sort. Meanwhile it is obvious
that if we split one sort into very many subsorts then we can formally
attain a
wrong result.

Let us suppose that the number of the heterogeneous centers is so great
 that
$h_i \gg 1$,
i.e. the fall of the supersaturation leads to the interruption  of
 formation of
the droplets.
We shall split this sort into so many subsorts that for every subsort
$h_{ij} \ll 1$,
i.e. we can see the exhaustion of the subsort in the separate process
of condensation.
Moreover, we can assume that this exhaustion finishes before
$z$ attains the quarter
of the length of the spectrum initiated by the fall of the supersaturation.
Really, as far as $\Delta_{ij} x$
is proportional to $\eta_{ij
\ tot}^{-1/4}$,
(where $\eta_{ij \ tot}$
is the total number of the centers of the given subsort) and $\delta_{ij} x $
doesn't depend on this quantity the required property is obvious.
Then after the summation over all subsorts we come to the conclusion that
the total number of the droplets formed up to $\Delta_i x / 4$
coincides with the total number of the heterogeneous centers.
This conclusion isn't valid.

The reason of the error is that the width of the spectrum is much more
small than the width in the process of the separate formation.
Evidently, the monodisperse
approximation doesn't work at such distances. Ordinary,
under the reasonable definition
of the heterogeneous sorts all characteristic lengths are different.
So, ordinary, there is no such effect in this situation.

Now we shall correct the theory. Note that the monodisperse approximation
leads to some already defined functional form for $g$
and for the supersaturation. We already know this form and now we need only
to establish the parameters in these functional dependencies.

We shall choose the unique length of the spectrum $\Delta x $.
At this very moment this value is unknown, but $\Delta x $
satisfies the following inequality:
$$
\Delta x  \leq \Delta_i x
$$
for every sort of the heterogeneous centers. Then for $g$
we have
\begin{equation}
\label{ggg}
          g_i(z) = \frac{N_i(\Delta x  / 4)}{n_{\infty}} z^3
\end{equation}
which leads to
\begin{equation}
N_{i(2)}(\Delta x / 4) = \eta_{i\ tot}(1 -  \theta_{i(2)}(\Delta x / 4))
\end{equation}
\begin{equation}
\label{ff}
         \theta_{i\ (2)} (\Delta x / 4)=
\exp[-f_{*\ i}
\frac{n_{\infty}}{\eta_{tot\ i}}
\int_0^z \exp(-\frac{\Gamma z^3}
{\Phi_* n_{\infty}} \sum_j N_j(\Delta x / 4)) dx ]
\end{equation}
We neglect here for simplicity the weak dependence $\Gamma_i = \Gamma$
on the sort of the centers, while the strong dependence $f_{* j}$
on the sort of centers
is taken into consideration. After the substitution we get the system
of the algebraic equations for $N_i(\Delta x / 4)$
\begin{equation}
\label{ghj}
          N_i (\Delta x / 4) =
\eta_{i\ tot}
(1 - \exp(-f_{*\ i}
\frac{n_{\infty}}{\eta_{tot\ i}}
\int_0^{\Delta x / 4}
\exp(-
\frac{\Gamma}{\Phi_*}
\frac{z^3}{n_{\infty}}
\sum_j N_j(\Delta x
/ 4)) dx))
\end{equation}

Let us simplify the last system. To calculate the integral one can note
that
\begin{equation}
                \int_0^x \exp(-x^3) dx \approx x \ \ \ \ x \leq 1/4
\end{equation}
So the trivial dependence of the r.h.s. on $N_j$
disappears. We come to
        \begin{equation}
\label{sf}
         N_i(\Delta x / 4)  = \eta_{i\ tot} ( 1 - \exp(-f_{*\ i}
\frac{n_{\infty}}{\eta_{tot\ i}}
\frac{\Delta
x }{4}))
\end{equation}

On the other hand we can utilize the sense of $\Delta x $
as the halfwidth of the spectrum due to the fall of the supersaturation:
\begin{equation}
\frac{\Delta x^3}{n_{\infty}} \sum_j
         N_i(\Delta x / 4)  \frac{\Gamma}{\Phi_*}
= 1
\end{equation}

After the substitution we can get the equation for
$\Delta x$
\begin{equation}
\frac{\Delta x^3}{n_{\infty}} \sum_j \eta_{j\ tot}
(1 - \exp( - f_{*\ j}
\frac{n_{\infty}}{\eta_{tot\ i}}
\frac{\Delta x}{4})) \frac{\Gamma}{\Phi_*}
= 1
\end{equation}
Equation (\ref{sf})
expresses $N_i(\Delta x / 4)$
through $\Delta x$
and solves the problem.

Equation (\ref{ggg})
gives the expression for $g$
and, thus,  for the supersaturation as the function of time.
The number of the heterogeneous centers can be found by (\ref{ff}).
The final values can be found by the  corresponding previous  formulas.

 We can estimate the relative error by the pseudo homogeneous case.
 Then we shall
obtain that the relative error can be roughly estimated\footnote{Precise
calculations for condensation of the separate sort give the relative
error of the monedisperse approximation as $0.0221$ in the total  number
of the droplets. The summation over all sorts can not increase the relative
error} as
\begin{equation} \label{93}
\frac{\mid N_{i}(\infty) - N_{final\ i}(\infty) \mid}{N_{i}(\infty)}
\leq \frac{\mid A -B \mid}{A} \sim 0.02
\end{equation}
The investigation of the process of the essential formation of the  droplets
 is completed.

\section{Concluding remarks}

As the result of the previous sections we know  the behavior of the
supersaturation and the behavior of the number of the free heterogeneous
centers during the period of the intensive formation of the droplets. It allows
to
get the main characteristic of the process of condensation - the total
number of the droplets formed on the heterogeneous centers of the different
nature.  The final formulas allow the interesting quantitative physical
analysis which is missed only due to the lack of the volume.  The evolution
of the already formed spectrum is much more simple. The high accuracy
in the determination of the supersaturation isn't necessary now. Due to
the proportionality of the velocity of growth to the supersaturation only
the high relative accuracy is necessary.  The description of the further
evolution is analogous to the  investigation of the homogeneous decay
made in \cite{9}.

The description of the further periods can be given with the help of
the direct application of the monodisperse approximation as in the case of
the homogeneous decay of the
metastable phase. We have the simple differential  equation
for some isolated hydrodynamic element
\begin{equation} \label{94}
\frac{\tau}{\alpha} \frac{dz}{dt} = \zeta =
\Phi_{*} - \sum_{i} N_{final\ i}(\infty)
\frac{z^3}{n_{\infty}}
\end{equation}
which can be easily integrated as far as
the r.h.s. doesn't depend on time.
All terms in this equation are known from the previous analysis. So
there are
no problems in the description of the process  until
the coalescence \cite{15},
 \cite{16}.

\pagebreak

\begin{picture}(400,500)
\put(51,438){.}
\put(51,435){.}
\put(52,433){.}
\put(52,430){.}
\put(53,428){.}
\put(54,426){.}
\put(54,423){.}
\put(55,421){.}
\put(55,419){.}
\put(56,416){.}
\put(57,414){.}
\put(57,412){.}
\put(58,410){.}
\put(58,407){.}
\put(59,405){.}
\put(60,403){.}
\put(60,401){.}
\put(61,398){.}
\put(61,396){.}
\put(62,394){.}
\put(63,392){.}
\put(63,389){.}
\put(64,387){.}
\put(64,385){.}
\put(65,383){.}
\put(66,381){.}
\put(66,379){.}
\put(67,376){.}
\put(67,374){.}
\put(68,372){.}
\put(69,370){.}
\put(69,368){.}
\put(70,366){.}
\put(70,364){.}
\put(71,362){.}
\put(72,360){.}
\put(72,358){.}
\put(73,356){.}
\put(73,354){.}
\put(74,351){.}
\put(75,349){.}
\put(75,347){.}
\put(76,345){.}
\put(76,343){.}
\put(77,341){.}
\put(78,339){.}
\put(78,337){.}
\put(79,336){.}
\put(79,334){.}
\put(80,332){.}
\put(81,330){.}
\put(81,328){.}
\put(82,326){.}
\put(82,324){.}
\put(83,322){.}
\put(84,320){.}
\put(84,318){.}
\put(85,316){.}
\put(85,314){.}
\put(86,313){.}
\put(87,311){.}
\put(87,309){.}
\put(88,307){.}
\put(88,305){.}
\put(89,303){.}
\put(90,302){.}
\put(90,300){.}
\put(91,298){.}
\put(91,296){.}
\put(92,294){.}
\put(93,293){.}
\put(93,291){.}
\put(94,289){.}
\put(94,287){.}
\put(95,286){.}
\put(96,284){.}
\put(96,282){.}
\put(97,280){.}
\put(97,279){.}
\put(98,277){.}
\put(99,275){.}
\put(99,274){.}
\put(100,272){.}
\put(100,270){.}
\put(101,269){.}
\put(102,267){.}
\put(102,265){.}
\put(103,264){.}
\put(103,262){.}
\put(104,261){.}
\put(105,259){.}
\put(105,257){.}
\put(106,256){.}
\put(106,254){.}
\put(107,253){.}
\put(108,251){.}
\put(108,249){.}
\put(109,248){.}
\put(109,246){.}
\put(110,245){.}
\put(111,243){.}
\put(111,242){.}
\put(112,240){.}
\put(112,239){.}
\put(113,237){.}
\put(114,236){.}
\put(114,234){.}
\put(115,233){.}
\put(115,231){.}
\put(116,230){.}
\put(117,228){.}
\put(117,227){.}
\put(118,225){.}
\put(118,224){.}
\put(119,223){.}
\put(120,221){.}
\put(120,220){.}
\put(121,218){.}
\put(121,217){.}
\put(122,216){.}
\put(123,214){.}
\put(123,213){.}
\put(124,211){.}
\put(124,210){.}
\put(125,209){.}
\put(126,207){.}
\put(126,206){.}
\put(127,205){.}
\put(127,203){.}
\put(128,202){.}
\put(129,201){.}
\put(129,199){.}
\put(130,198){.}
\put(130,197){.}
\put(131,196){.}
\put(132,194){.}
\put(132,193){.}
\put(133,192){.}
\put(133,191){.}
\put(134,189){.}
\put(135,188){.}
\put(135,187){.}
\put(136,186){.}
\put(136,184){.}
\put(137,183){.}
\put(138,182){.}
\put(138,181){.}
\put(139,180){.}
\put(139,178){.}
\put(140,177){.}
\put(141,176){.}
\put(141,175){.}
\put(142,174){.}
\put(142,173){.}
\put(143,171){.}
\put(144,170){.}
\put(144,169){.}
\put(145,168){.}
\put(145,167){.}
\put(146,166){.}
\put(147,165){.}
\put(147,164){.}
\put(148,162){.}
\put(148,161){.}
\put(149,160){.}
\put(150,159){.}
\put(150,158){.}
\put(151,157){.}
\put(151,156){.}
\put(152,155){.}
\put(153,154){.}
\put(153,153){.}
\put(154,152){.}
\put(154,151){.}
\put(155,150){.}
\put(156,149){.}
\put(156,148){.}
\put(157,147){.}
\put(157,146){.}
\put(158,145){.}
\put(159,144){.}
\put(159,143){.}
\put(160,142){.}
\put(160,141){.}
\put(161,140){.}
\put(162,139){.}
\put(162,138){.}
\put(163,137){.}
\put(163,136){.}
\put(164,135){.}
\put(165,134){.}
\put(165,133){.}
\put(166,133){.}
\put(166,132){.}
\put(167,131){.}
\put(168,130){.}
\put(168,129){.}
\put(169,128){.}
\put(169,127){.}
\put(170,126){.}
\put(171,126){.}
\put(171,125){.}
\put(172,124){.}
\put(172,123){.}
\put(173,122){.}
\put(174,121){.}
\put(174,120){.}
\put(175,120){.}
\put(175,119){.}
\put(176,118){.}
\put(177,117){.}
\put(177,116){.}
\put(178,116){.}
\put(178,115){.}
\put(179,114){.}
\put(180,113){.}
\put(180,113){.}
\put(181,112){.}
\put(181,111){.}
\put(182,110){.}
\put(183,109){.}
\put(183,109){.}
\put(184,108){.}
\put(184,107){.}
\put(185,107){.}
\put(186,106){.}
\put(186,105){.}
\put(187,104){.}
\put(187,104){.}
\put(188,103){.}
\put(189,102){.}
\put(189,102){.}
\put(190,101){.}
\put(190,100){.}
\put(191,100){.}
\put(192,99){.}
\put(192,98){.}
\put(193,98){.}
\put(193,97){.}
\put(194,96){.}
\put(195,96){.}
\put(195,95){.}
\put(196,94){.}
\put(196,94){.}
\put(197,93){.}
\put(198,92){.}
\put(198,92){.}
\put(199,91){.}
\put(199,91){.}
\put(200,90){.}
\put(201,89){.}
\put(201,89){.}
\put(202,88){.}
\put(202,88){.}
\put(203,87){.}
\put(204,86){.}
\put(204,86){.}
\put(205,85){.}
\put(205,85){.}
\put(206,84){.}
\put(207,84){.}
\put(207,83){.}
\put(208,83){.}
\put(208,82){.}
\put(209,82){.}
\put(210,81){.}
\put(210,80){.}
\put(211,80){.}
\put(211,79){.}
\put(212,79){.}
\put(213,78){.}
\put(213,78){.}
\put(214,77){.}
\put(214,77){.}
\put(215,76){.}
\put(216,76){.}
\put(216,75){.}
\put(217,75){.}
\put(217,75){.}
\put(218,74){.}
\put(219,74){.}
\put(219,73){.}
\put(220,73){.}
\put(220,72){.}
\put(221,72){.}
\put(222,71){.}
\put(222,71){.}
\put(223,70){.}
\put(223,70){.}
\put(224,70){.}
\put(225,69){.}
\put(225,69){.}
\put(226,68){.}
\put(226,68){.}
\put(227,68){.}
\put(228,67){.}
\put(228,67){.}
\put(229,66){.}
\put(229,66){.}
\put(230,66){.}
\put(231,65){.}
\put(231,65){.}
\put(232,64){.}
\put(232,64){.}
\put(233,64){.}
\put(234,63){.}
\put(234,63){.}
\put(235,63){.}
\put(235,62){.}
\put(236,62){.}
\put(237,62){.}
\put(237,61){.}
\put(238,61){.}
\put(238,61){.}
\put(239,60){.}
\put(240,60){.}
\put(240,60){.}
\put(241,59){.}
\put(241,59){.}
\put(242,59){.}
\put(243,58){.}
\put(243,58){.}
\put(244,58){.}
\put(244,57){.}
\put(245,57){.}
\put(246,57){.}
\put(246,57){.}
\put(247,56){.}
\put(247,56){.}
\put(248,56){.}
\put(249,55){.}
\put(249,55){.}
\put(250,55){.}
\put(250,55){.}
\put(251,54){.}
\put(252,54){.}
\put(252,54){.}
\put(253,54){.}
\put(253,53){.}
\put(254,53){.}
\put(255,53){.}
\put(255,53){.}
\put(256,52){.}
\put(256,52){.}
\put(257,52){.}
\put(258,52){.}
\put(258,51){.}
\put(259,51){.}
\put(259,51){.}
\put(260,51){.}
\put(261,51){.}
\put(261,50){.}
\put(262,50){.}
\put(262,50){.}
\put(263,50){.}
\put(264,50){.}
\put(264,49){.}
\put(265,49){.}
\put(265,49){.}
\put(266,49){.}
\put(267,49){.}
\put(267,48){.}
\put(268,48){.}
\put(268,48){.}
\put(269,48){.}
\put(270,48){.}
\put(270,48){.}
\put(271,47){.}
\put(271,47){.}
\put(272,47){.}
\put(273,47){.}
\put(273,47){.}
\put(274,47){.}
\put(274,46){.}
\put(275,46){.}
\put(276,46){.}
\put(276,46){.}
\put(277,46){.}
\put(277,46){.}
\put(278,46){.}
\put(279,45){.}
\put(279,45){.}
\put(280,45){.}
\put(280,45){.}
\put(281,45){.}
\put(282,45){.}
\put(282,45){.}
\put(283,44){.}
\put(283,44){.}
\put(284,44){.}
\put(285,44){.}
\put(285,44){.}
\put(286,44){.}
\put(286,44){.}
\put(287,44){.}
\put(288,44){.}
\put(288,43){.}
\put(289,43){.}
\put(289,43){.}
\put(290,43){.}
\put(291,43){.}
\put(291,43){.}
\put(292,43){.}
\put(292,43){.}
\put(293,43){.}
\put(294,43){.}
\put(294,43){.}
\put(295,42){.}
\put(295,42){.}
\put(296,42){.}
\put(297,42){.}
\put(297,42){.}
\put(298,42){.}
\put(298,42){.}
\put(299,42){.}
\put(300,42){.}
\put(300,42){.}
\put(301,42){.}
\put(301,42){.}
\put(302,42){.}
\put(303,42){.}
\put(303,42){.}
\put(304,41){.}
\put(304,41){.}
\put(305,41){.}
\put(306,41){.}
\put(306,41){.}
\put(307,41){.}
\put(307,41){.}
\put(308,41){.}
\put(309,41){.}
\put(309,41){.}
\put(310,41){.}
\put(310,41){.}
\put(311,41){.}
\put(312,41){.}
\put(312,41){.}
\put(313,41){.}
\put(313,41){.}
\put(314,41){.}
\put(315,41){.}
\put(315,41){.}
\put(316,41){.}
\put(316,41){.}
\put(317,41){.}
\put(318,41){.}
\put(318,40){.}
\put(319,40){.}
\put(319,40){.}
\put(320,40){.}
\put(321,40){.}
\put(321,40){.}
\put(322,40){.}
\put(322,40){.}
\put(323,40){.}
\put(324,40){.}
\put(324,40){.}
\put(325,40){.}
\put(325,40){.}
\put(326,40){.}
\put(327,40){.}
\put(327,40){.}
\put(328,40){.}
\put(328,40){.}
\put(329,40){.}
\put(330,40){.}
\put(330,40){.}
\put(331,40){.}
\put(331,40){.}
\put(332,40){.}
\put(333,40){.}
\put(333,40){.}
\put(334,40){.}
\put(334,40){.}
\put(335,40){.}
\put(336,40){.}
\put(336,40){.}
\put(337,40){.}
\put(337,40){.}
\put(338,40){.}
\put(339,40){.}
\put(339,40){.}
\put(340,40){.}
\put(340,40){.}
\put(341,40){.}
\put(342,40){.}
\put(342,40){.}
\put(343,40){.}
\put(343,40){.}
\put(344,40){.}
\put(345,40){.}
\put(345,40){.}
\put(346,40){.}
\put(346,40){.}
\put(347,40){.}
\put(348,40){.}
\put(348,40){.}
\put(349,40){.}
\put(349,40){.}
\put(350,40){.}
\put(51,438){.}
\put(51,435){.}
\put(52,433){.}
\put(52,430){.}
\put(53,428){.}
\put(54,426){.}
\put(54,423){.}
\put(55,421){.}
\put(55,419){.}
\put(56,416){.}
\put(57,414){.}
\put(57,412){.}
\put(58,410){.}
\put(58,407){.}
\put(59,405){.}
\put(60,403){.}
\put(60,401){.}
\put(61,398){.}
\put(61,396){.}
\put(62,394){.}
\put(63,392){.}
\put(63,389){.}
\put(64,387){.}
\put(64,385){.}
\put(65,383){.}
\put(66,381){.}
\put(66,379){.}
\put(67,376){.}
\put(67,374){.}
\put(68,372){.}
\put(69,370){.}
\put(69,368){.}
\put(70,366){.}
\put(70,364){.}
\put(71,362){.}
\put(72,360){.}
\put(72,358){.}
\put(73,356){.}
\put(73,353){.}
\put(74,351){.}
\put(75,349){.}
\put(75,347){.}
\put(76,345){.}
\put(76,343){.}
\put(77,341){.}
\put(78,339){.}
\put(78,337){.}
\put(79,335){.}
\put(79,334){.}
\put(80,332){.}
\put(81,330){.}
\put(81,328){.}
\put(82,326){.}
\put(82,324){.}
\put(83,322){.}
\put(84,320){.}
\put(84,318){.}
\put(85,316){.}
\put(85,314){.}
\put(86,313){.}
\put(87,311){.}
\put(87,309){.}
\put(88,307){.}
\put(88,305){.}
\put(89,303){.}
\put(90,302){.}
\put(90,300){.}
\put(91,298){.}
\put(91,296){.}
\put(92,294){.}
\put(93,293){.}
\put(93,291){.}
\put(94,289){.}
\put(94,287){.}
\put(95,286){.}
\put(96,284){.}
\put(96,282){.}
\put(97,280){.}
\put(97,279){.}
\put(98,277){.}
\put(99,275){.}
\put(99,274){.}
\put(100,272){.}
\put(100,270){.}
\put(101,269){.}
\put(102,267){.}
\put(102,265){.}
\put(103,264){.}
\put(103,262){.}
\put(104,260){.}
\put(105,259){.}
\put(105,257){.}
\put(106,255){.}
\put(106,254){.}
\put(107,252){.}
\put(108,251){.}
\put(108,249){.}
\put(109,248){.}
\put(109,246){.}
\put(110,244){.}
\put(111,243){.}
\put(111,241){.}
\put(112,240){.}
\put(112,238){.}
\put(113,237){.}
\put(114,235){.}
\put(114,234){.}
\put(115,232){.}
\put(115,231){.}
\put(116,229){.}
\put(117,228){.}
\put(117,226){.}
\put(118,225){.}
\put(118,224){.}
\put(119,222){.}
\put(120,221){.}
\put(120,219){.}
\put(121,218){.}
\put(121,216){.}
\put(122,215){.}
\put(123,214){.}
\put(123,212){.}
\put(124,211){.}
\put(124,209){.}
\put(125,208){.}
\put(126,207){.}
\put(126,205){.}
\put(127,204){.}
\put(127,203){.}
\put(128,201){.}
\put(129,200){.}
\put(129,199){.}
\put(130,197){.}
\put(130,196){.}
\put(131,195){.}
\put(132,193){.}
\put(132,192){.}
\put(133,191){.}
\put(133,190){.}
\put(134,188){.}
\put(135,187){.}
\put(135,186){.}
\put(136,185){.}
\put(136,183){.}
\put(137,182){.}
\put(138,181){.}
\put(138,180){.}
\put(139,178){.}
\put(139,177){.}
\put(140,176){.}
\put(141,175){.}
\put(141,174){.}
\put(142,173){.}
\put(142,171){.}
\put(143,170){.}
\put(144,169){.}
\put(144,168){.}
\put(145,167){.}
\put(145,166){.}
\put(146,164){.}
\put(147,163){.}
\put(147,162){.}
\put(148,161){.}
\put(148,160){.}
\put(149,159){.}
\put(150,158){.}
\put(150,157){.}
\put(151,156){.}
\put(151,155){.}
\put(152,153){.}
\put(153,152){.}
\put(153,151){.}
\put(154,150){.}
\put(154,149){.}
\put(155,148){.}
\put(156,147){.}
\put(156,146){.}
\put(157,145){.}
\put(157,144){.}
\put(158,143){.}
\put(159,142){.}
\put(159,141){.}
\put(160,140){.}
\put(160,139){.}
\put(161,138){.}
\put(162,137){.}
\put(162,136){.}
\put(163,135){.}
\put(163,134){.}
\put(164,133){.}
\put(165,132){.}
\put(165,131){.}
\put(166,131){.}
\put(166,130){.}
\put(167,129){.}
\put(168,128){.}
\put(168,127){.}
\put(169,126){.}
\put(169,125){.}
\put(170,124){.}
\put(171,123){.}
\put(171,122){.}
\put(172,122){.}
\put(172,121){.}
\put(173,120){.}
\put(174,119){.}
\put(174,118){.}
\put(175,117){.}
\put(175,116){.}
\put(176,116){.}
\put(177,115){.}
\put(177,114){.}
\put(178,113){.}
\put(178,112){.}
\put(179,112){.}
\put(180,111){.}
\put(180,110){.}
\put(181,109){.}
\put(181,108){.}
\put(182,108){.}
\put(183,107){.}
\put(183,106){.}
\put(184,105){.}
\put(184,105){.}
\put(185,104){.}
\put(186,103){.}
\put(186,102){.}
\put(187,102){.}
\put(187,101){.}
\put(188,100){.}
\put(189,100){.}
\put(189,99){.}
\put(190,98){.}
\put(190,97){.}
\put(191,97){.}
\put(192,96){.}
\put(192,95){.}
\put(193,95){.}
\put(193,94){.}
\put(194,93){.}
\put(195,93){.}
\put(195,92){.}
\put(196,91){.}
\put(196,91){.}
\put(197,90){.}
\put(198,89){.}
\put(198,89){.}
\put(199,88){.}
\put(199,88){.}
\put(200,87){.}
\put(201,86){.}
\put(201,86){.}
\put(202,85){.}
\put(202,85){.}
\put(203,84){.}
\put(204,83){.}
\put(204,83){.}
\put(205,82){.}
\put(205,82){.}
\put(206,81){.}
\put(207,81){.}
\put(207,80){.}
\put(208,79){.}
\put(208,79){.}
\put(209,78){.}
\put(210,78){.}
\put(210,77){.}
\put(211,77){.}
\put(211,76){.}
\put(212,76){.}
\put(213,75){.}
\put(213,75){.}
\put(214,74){.}
\put(214,74){.}
\put(215,73){.}
\put(216,73){.}
\put(216,72){.}
\put(217,72){.}
\put(217,71){.}
\put(218,71){.}
\put(219,70){.}
\put(219,70){.}
\put(220,70){.}
\put(220,69){.}
\put(221,69){.}
\put(222,68){.}
\put(222,68){.}
\put(223,67){.}
\put(223,67){.}
\put(224,66){.}
\put(225,66){.}
\put(225,66){.}
\put(226,65){.}
\put(226,65){.}
\put(227,64){.}
\put(228,64){.}
\put(228,64){.}
\put(229,63){.}
\put(229,63){.}
\put(230,62){.}
\put(231,62){.}
\put(231,62){.}
\put(232,61){.}
\put(232,61){.}
\put(233,61){.}
\put(234,60){.}
\put(234,60){.}
\put(235,60){.}
\put(235,59){.}
\put(236,59){.}
\put(237,59){.}
\put(237,58){.}
\put(238,58){.}
\put(238,58){.}
\put(239,57){.}
\put(240,57){.}
\put(240,57){.}
\put(241,56){.}
\put(241,56){.}
\put(242,56){.}
\put(243,55){.}
\put(243,55){.}
\put(244,55){.}
\put(244,55){.}
\put(245,54){.}
\put(246,54){.}
\put(246,54){.}
\put(247,54){.}
\put(247,53){.}
\put(248,53){.}
\put(249,53){.}
\put(249,52){.}
\put(250,52){.}
\put(250,52){.}
\put(251,52){.}
\put(252,52){.}
\put(252,51){.}
\put(253,51){.}
\put(253,51){.}
\put(254,51){.}
\put(255,50){.}
\put(255,50){.}
\put(256,50){.}
\put(256,50){.}
\put(257,49){.}
\put(258,49){.}
\put(258,49){.}
\put(259,49){.}
\put(259,49){.}
\put(260,48){.}
\put(261,48){.}
\put(261,48){.}
\put(262,48){.}
\put(262,48){.}
\put(263,48){.}
\put(264,47){.}
\put(264,47){.}
\put(265,47){.}
\put(265,47){.}
\put(266,47){.}
\put(267,47){.}
\put(267,46){.}
\put(268,46){.}
\put(268,46){.}
\put(269,46){.}
\put(270,46){.}
\put(270,46){.}
\put(271,45){.}
\put(271,45){.}
\put(272,45){.}
\put(273,45){.}
\put(273,45){.}
\put(274,45){.}
\put(274,45){.}
\put(275,45){.}
\put(276,44){.}
\put(276,44){.}
\put(277,44){.}
\put(277,44){.}
\put(278,44){.}
\put(279,44){.}
\put(279,44){.}
\put(280,44){.}
\put(280,44){.}
\put(281,43){.}
\put(282,43){.}
\put(282,43){.}
\put(283,43){.}
\put(283,43){.}
\put(284,43){.}
\put(285,43){.}
\put(285,43){.}
\put(286,43){.}
\put(286,43){.}
\put(287,43){.}
\put(288,42){.}
\put(288,42){.}
\put(289,42){.}
\put(289,42){.}
\put(290,42){.}
\put(291,42){.}
\put(291,42){.}
\put(292,42){.}
\put(292,42){.}
\put(293,42){.}
\put(294,42){.}
\put(294,42){.}
\put(295,42){.}
\put(295,42){.}
\put(296,41){.}
\put(297,41){.}
\put(297,41){.}
\put(298,41){.}
\put(298,41){.}
\put(299,41){.}
\put(300,41){.}
\put(300,41){.}
\put(301,41){.}
\put(301,41){.}
\put(302,41){.}
\put(303,41){.}
\put(303,41){.}
\put(304,41){.}
\put(304,41){.}
\put(305,41){.}
\put(306,41){.}
\put(306,41){.}
\put(307,41){.}
\put(307,41){.}
\put(308,41){.}
\put(309,41){.}
\put(309,41){.}
\put(310,41){.}
\put(310,41){.}
\put(311,40){.}
\put(312,40){.}
\put(312,40){.}
\put(313,40){.}
\put(313,40){.}
\put(314,40){.}
\put(315,40){.}
\put(315,40){.}
\put(316,40){.}
\put(316,40){.}
\put(317,40){.}
\put(318,40){.}
\put(318,40){.}
\put(319,40){.}
\put(319,40){.}
\put(320,40){.}
\put(321,40){.}
\put(321,40){.}
\put(322,40){.}
\put(322,40){.}
\put(323,40){.}
\put(324,40){.}
\put(324,40){.}
\put(325,40){.}
\put(325,40){.}
\put(326,40){.}
\put(327,40){.}
\put(327,40){.}
\put(328,40){.}
\put(328,40){.}
\put(329,40){.}
\put(330,40){.}
\put(330,40){.}
\put(331,40){.}
\put(331,40){.}
\put(332,40){.}
\put(333,40){.}
\put(333,40){.}
\put(334,40){.}
\put(334,40){.}
\put(335,40){.}
\put(336,40){.}
\put(336,40){.}
\put(337,40){.}
\put(337,40){.}
\put(338,40){.}
\put(339,40){.}
\put(339,40){.}
\put(340,40){.}
\put(340,40){.}
\put(341,40){.}
\put(342,40){.}
\put(342,40){.}
\put(343,40){.}
\put(343,40){.}
\put(344,40){.}
\put(345,40){.}
\put(345,40){.}
\put(346,40){.}
\put(346,40){.}
\put(347,40){.}
\put(348,40){.}
\put(348,40){.}
\put(349,40){.}
\put(349,40){.}
\put(350,40){.}
\put(125,40){\line(0,-1){5}}
\put(125,20){$\Delta_{1/2} x$}
\put(112,40){\line(0,1){5}}
\put(100,50){$\delta_{1/2} x$}
\put(25,40){\vector(1,0){350}}
\put(50,40){\line(0,1){440}}
\put(50,25){0}
\put(350,25){$z$}
\put(350,40){\line(0,1){5}}
\put(370,25){$x$}
\put(200,60){$s_{bel}$}
\put(230,70){$s_{ab}$}
\put(170,3){$Figure \ 1$}
\end{picture}

$$ Behavior \ of \ functions \ s_{bel} \ and \ s_{ab} . $$

\pagebreak

\begin{picture}(400,250)
\put(51,240){.}
\put(52,240){.}
\put(54,240){.}
\put(55,240){.}
\put(56,240){.}
\put(57,240){.}
\put(58,240){.}
\put(60,240){.}
\put(61,240){.}
\put(62,240){.}
\put(63,240){.}
\put(64,240){.}
\put(66,240){.}
\put(67,240){.}
\put(68,240){.}
\put(69,240){.}
\put(70,240){.}
\put(72,240){.}
\put(73,240){.}
\put(74,240){.}
\put(75,240){.}
\put(76,240){.}
\put(78,240){.}
\put(79,240){.}
\put(80,240){.}
\put(81,240){.}
\put(82,239){.}
\put(84,239){.}
\put(85,239){.}
\put(86,239){.}
\put(87,239){.}
\put(88,239){.}
\put(90,239){.}
\put(91,239){.}
\put(92,238){.}
\put(93,238){.}
\put(94,238){.}
\put(96,238){.}
\put(97,238){.}
\put(98,237){.}
\put(99,237){.}
\put(100,237){.}
\put(102,236){.}
\put(103,236){.}
\put(104,236){.}
\put(105,235){.}
\put(106,235){.}
\put(108,235){.}
\put(109,234){.}
\put(110,234){.}
\put(111,233){.}
\put(112,233){.}
\put(114,232){.}
\put(115,231){.}
\put(116,231){.}
\put(117,230){.}
\put(118,229){.}
\put(120,229){.}
\put(121,228){.}
\put(122,227){.}
\put(123,226){.}
\put(124,225){.}
\put(126,224){.}
\put(127,223){.}
\put(128,222){.}
\put(129,221){.}
\put(130,220){.}
\put(132,219){.}
\put(133,218){.}
\put(134,217){.}
\put(135,215){.}
\put(136,214){.}
\put(138,213){.}
\put(139,211){.}
\put(140,210){.}
\put(141,208){.}
\put(142,207){.}
\put(144,205){.}
\put(145,203){.}
\put(146,202){.}
\put(147,200){.}
\put(148,198){.}
\put(150,196){.}
\put(151,195){.}
\put(152,193){.}
\put(153,191){.}
\put(154,189){.}
\put(156,187){.}
\put(157,184){.}
\put(158,182){.}
\put(159,180){.}
\put(160,178){.}
\put(162,176){.}
\put(163,173){.}
\put(164,171){.}
\put(165,169){.}
\put(166,166){.}
\put(168,164){.}
\put(169,162){.}
\put(170,159){.}
\put(171,157){.}
\put(172,154){.}
\put(174,152){.}
\put(175,149){.}
\put(176,147){.}
\put(177,144){.}
\put(178,141){.}
\put(180,139){.}
\put(181,136){.}
\put(182,134){.}
\put(183,131){.}
\put(184,128){.}
\put(186,126){.}
\put(187,123){.}
\put(188,121){.}
\put(189,118){.}
\put(190,116){.}
\put(192,113){.}
\put(193,111){.}
\put(194,108){.}
\put(195,106){.}
\put(196,103){.}
\put(198,101){.}
\put(199,99){.}
\put(200,96){.}
\put(201,94){.}
\put(202,92){.}
\put(204,90){.}
\put(205,88){.}
\put(206,86){.}
\put(207,83){.}
\put(208,81){.}
\put(210,79){.}
\put(211,78){.}
\put(212,76){.}
\put(213,74){.}
\put(214,72){.}
\put(216,71){.}
\put(217,69){.}
\put(218,67){.}
\put(219,66){.}
\put(220,64){.}
\put(222,63){.}
\put(223,62){.}
\put(224,60){.}
\put(225,59){.}
\put(226,58){.}
\put(228,57){.}
\put(229,56){.}
\put(230,54){.}
\put(231,54){.}
\put(232,53){.}
\put(234,52){.}
\put(235,51){.}
\put(236,50){.}
\put(237,49){.}
\put(238,49){.}
\put(240,48){.}
\put(241,47){.}
\put(242,47){.}
\put(243,46){.}
\put(244,46){.}
\put(246,45){.}
\put(247,45){.}
\put(248,44){.}
\put(249,44){.}
\put(250,44){.}
\put(252,43){.}
\put(253,43){.}
\put(254,43){.}
\put(255,42){.}
\put(256,42){.}
\put(258,42){.}
\put(259,42){.}
\put(260,42){.}
\put(261,41){.}
\put(262,41){.}
\put(264,41){.}
\put(265,41){.}
\put(266,41){.}
\put(267,41){.}
\put(268,41){.}
\put(270,41){.}
\put(271,41){.}
\put(272,40){.}
\put(273,40){.}
\put(274,40){.}
\put(276,40){.}
\put(277,40){.}
\put(278,40){.}
\put(279,40){.}
\put(280,40){.}
\put(282,40){.}
\put(283,40){.}
\put(284,40){.}
\put(285,40){.}
\put(286,40){.}
\put(288,40){.}
\put(289,40){.}
\put(290,40){.}
\put(291,40){.}
\put(292,40){.}
\put(294,40){.}
\put(295,40){.}
\put(296,40){.}
\put(297,40){.}
\put(298,40){.}
\put(300,40){.}
\put(301,40){.}
\put(302,40){.}
\put(303,40){.}
\put(304,40){.}
\put(306,40){.}
\put(307,40){.}
\put(308,40){.}
\put(309,40){.}
\put(310,40){.}
\put(312,40){.}
\put(313,40){.}
\put(314,40){.}
\put(315,40){.}
\put(316,40){.}
\put(318,40){.}
\put(319,40){.}
\put(320,40){.}
\put(321,40){.}
\put(322,40){.}
\put(324,40){.}
\put(325,40){.}
\put(326,40){.}
\put(327,40){.}
\put(328,40){.}
\put(330,40){.}
\put(331,40){.}
\put(332,40){.}
\put(333,40){.}
\put(334,40){.}
\put(336,40){.}
\put(337,40){.}
\put(338,40){.}
\put(339,40){.}
\put(340,40){.}
\put(342,40){.}
\put(343,40){.}
\put(344,40){.}
\put(345,40){.}
\put(346,40){.}
\put(348,40){.}
\put(349,40){.}
\put(350,40){.}
\put(51,240){.}
\put(52,240){.}
\put(54,240){.}
\put(55,240){.}
\put(56,240){.}
\put(57,240){.}
\put(58,240){.}
\put(60,240){.}
\put(61,240){.}
\put(62,240){.}
\put(63,240){.}
\put(64,240){.}
\put(66,240){.}
\put(67,240){.}
\put(68,240){.}
\put(69,240){.}
\put(70,239){.}
\put(72,239){.}
\put(73,239){.}
\put(74,239){.}
\put(75,239){.}
\put(76,239){.}
\put(78,239){.}
\put(79,238){.}
\put(80,238){.}
\put(81,238){.}
\put(82,238){.}
\put(84,237){.}
\put(85,237){.}
\put(86,237){.}
\put(87,236){.}
\put(88,236){.}
\put(90,236){.}
\put(91,235){.}
\put(92,235){.}
\put(93,234){.}
\put(94,234){.}
\put(96,233){.}
\put(97,233){.}
\put(98,232){.}
\put(99,232){.}
\put(100,231){.}
\put(102,231){.}
\put(103,230){.}
\put(104,229){.}
\put(105,228){.}
\put(106,228){.}
\put(108,227){.}
\put(109,226){.}
\put(110,225){.}
\put(111,224){.}
\put(112,224){.}
\put(114,223){.}
\put(115,222){.}
\put(116,221){.}
\put(117,220){.}
\put(118,219){.}
\put(120,218){.}
\put(121,216){.}
\put(122,215){.}
\put(123,214){.}
\put(124,213){.}
\put(126,212){.}
\put(127,210){.}
\put(128,209){.}
\put(129,208){.}
\put(130,206){.}
\put(132,205){.}
\put(133,204){.}
\put(134,202){.}
\put(135,201){.}
\put(136,199){.}
\put(138,198){.}
\put(139,196){.}
\put(140,195){.}
\put(141,193){.}
\put(142,191){.}
\put(144,190){.}
\put(145,188){.}
\put(146,186){.}
\put(147,185){.}
\put(148,183){.}
\put(150,181){.}
\put(151,179){.}
\put(152,177){.}
\put(153,176){.}
\put(154,174){.}
\put(156,172){.}
\put(157,170){.}
\put(158,168){.}
\put(159,166){.}
\put(160,164){.}
\put(162,162){.}
\put(163,160){.}
\put(164,158){.}
\put(165,156){.}
\put(166,155){.}
\put(168,153){.}
\put(169,151){.}
\put(170,149){.}
\put(171,147){.}
\put(172,145){.}
\put(174,143){.}
\put(175,141){.}
\put(176,139){.}
\put(177,137){.}
\put(178,135){.}
\put(180,133){.}
\put(181,131){.}
\put(182,129){.}
\put(183,127){.}
\put(184,125){.}
\put(186,123){.}
\put(187,121){.}
\put(188,119){.}
\put(189,117){.}
\put(190,115){.}
\put(192,113){.}
\put(193,111){.}
\put(194,110){.}
\put(195,108){.}
\put(196,106){.}
\put(198,104){.}
\put(199,102){.}
\put(200,101){.}
\put(201,99){.}
\put(202,97){.}
\put(204,96){.}
\put(205,94){.}
\put(206,92){.}
\put(207,91){.}
\put(208,89){.}
\put(210,88){.}
\put(211,86){.}
\put(212,84){.}
\put(213,83){.}
\put(214,82){.}
\put(216,80){.}
\put(217,79){.}
\put(218,77){.}
\put(219,76){.}
\put(220,75){.}
\put(222,74){.}
\put(223,72){.}
\put(224,71){.}
\put(225,70){.}
\put(226,69){.}
\put(228,68){.}
\put(229,67){.}
\put(230,65){.}
\put(231,64){.}
\put(232,63){.}
\put(234,62){.}
\put(235,61){.}
\put(236,61){.}
\put(237,60){.}
\put(238,59){.}
\put(240,58){.}
\put(241,57){.}
\put(242,56){.}
\put(243,56){.}
\put(244,55){.}
\put(246,54){.}
\put(247,54){.}
\put(248,53){.}
\put(249,52){.}
\put(250,52){.}
\put(252,51){.}
\put(253,50){.}
\put(254,50){.}
\put(255,49){.}
\put(256,49){.}
\put(258,48){.}
\put(259,48){.}
\put(260,48){.}
\put(261,47){.}
\put(262,47){.}
\put(264,46){.}
\put(265,46){.}
\put(266,46){.}
\put(267,45){.}
\put(268,45){.}
\put(270,45){.}
\put(271,44){.}
\put(272,44){.}
\put(273,44){.}
\put(274,44){.}
\put(276,43){.}
\put(277,43){.}
\put(278,43){.}
\put(279,43){.}
\put(280,43){.}
\put(282,42){.}
\put(283,42){.}
\put(284,42){.}
\put(285,42){.}
\put(286,42){.}
\put(288,42){.}
\put(289,42){.}
\put(290,42){.}
\put(291,41){.}
\put(292,41){.}
\put(294,41){.}
\put(295,41){.}
\put(296,41){.}
\put(297,41){.}
\put(298,41){.}
\put(300,41){.}
\put(301,41){.}
\put(302,41){.}
\put(303,41){.}
\put(304,41){.}
\put(306,41){.}
\put(307,41){.}
\put(308,40){.}
\put(309,40){.}
\put(310,40){.}
\put(312,40){.}
\put(313,40){.}
\put(314,40){.}
\put(315,40){.}
\put(316,40){.}
\put(318,40){.}
\put(319,40){.}
\put(320,40){.}
\put(321,40){.}
\put(322,40){.}
\put(324,40){.}
\put(325,40){.}
\put(326,40){.}
\put(327,40){.}
\put(328,40){.}
\put(330,40){.}
\put(331,40){.}
\put(332,40){.}
\put(333,40){.}
\put(334,40){.}
\put(336,40){.}
\put(337,40){.}
\put(338,40){.}
\put(339,40){.}
\put(340,40){.}
\put(342,40){.}
\put(343,40){.}
\put(344,40){.}
\put(345,40){.}
\put(346,40){.}
\put(348,40){.}
\put(349,40){.}
\put(350,40){.}
\put(25,40){\vector(1,0){350}}
\put(50,40){\line(0,1){205}}
\put(50,25){0}
\put(370,25){$x$}
\put(200,60){$f_{1}$}
\put(230,70){$f_{appr}$}
\put(170,10){$lFigure \ 2 $}
\end{picture}

$$Form \ of \ the \ spectrum \  in \ the \ "monodisperse \ approximation".
$$

\end{document}